\documentclass[12pt]{article}

\usepackage[letterpaper, left=0.8in, right=0.8in, top=1in, bottom=1in]{geometry}
\usepackage{amsmath, amssymb, amsthm, bm}
\usepackage{graphicx} 
\usepackage{booktabs} 
\usepackage[round,authoryear]{natbib}
\usepackage{hyperref} 
\usepackage{setspace} 

\usepackage{subcaption}
\usepackage{float}   

\usepackage{amsthm}

\usepackage{algorithm}
\usepackage{algpseudocode}
\usepackage{adjustbox}
\usepackage{xcolor}

\usepackage{mathtools}

\newtheorem{proposition}{Proposition}

\newcommand{\new}{\mathrm{new}}
\newcommand{\E}{\mathbb{E}}
\newcommand{\Var}{\mathrm{Var}}
\newcommand{\Cov}{\mathrm{Cov}}

\usepackage{authblk}
\author[1]{Ying Zhang\thanks{zhang8yg@mail.uc.edu}}
\author[1]{Emily L. Kang\thanks{kangel@ucmail.uc.edu}}
\author[2]{Shuo Li\thanks{li.s.31@pg.com}}

\affil[1]{Division of Statistics and Data Science, Department of Mathematical Sciences, University of Cincinnati, Cincinnati, Ohio, U.S.A.}
\affil[2]{Research \& Development, Fabric \& Home Care Sector, 
The Procter \& Gamble Company, Cincinnati, Ohio, U.S.A.
}

\title{Efficient Gaussian Process Modeling for Large Replicated and Censored Data in Industrial Research}
\date{}

\begin{document}

\maketitle
\vspace{-3.4em} 
{\setlength{\parskip}{0pt}
\begin{abstract}
Gaussian process regression is widely used for modeling responses in both physical and computer experiments. In practice, data from physical experiments may include replications and be subject to censoring due to equipment limitations or contextual constraints\textcolor{blue}{, and such datasets can be large, posing a scalability challenge for standard Gaussian process inference}. Motivated by a real-world industrial research study, we develop a Gaussian process modeling framework that efficiently analyzes large replicated and censored data. Our approach integrates a Vecchia approximation for computational scalability with a likelihood-based formulation that fully exploits replication and rigorously incorporates information from censored observations, analogous to exact simulation from truncated distributions. Through experiments on synthetic data and a real case study from industrial research, we demonstrate that the proposed method yields improved predictive accuracy and reliable uncertainty quantification compared to existing approaches, making it well suited for large-scale applications with structured replication and incomplete data.
\end{abstract}
\noindent
\begin{minipage}{\textwidth}
\end{minipage}
\par}

\newpage

\section{Introduction}

In manufacturing and engineering research, experiments are routinely conducted to study the functional relationship between a set of input variables and an output of interest. To account for measurement variability, it is common practice to record multiple responses at the same input location, resulting in replicated observations. At the same time, data collection is often subject to practical limitations. Measurements may be censored due to detection limits of measuring instruments or operational constraints, leading to partially observed responses. As a result, real-world datasets frequently exhibit both replication and censoring, particularly in large-scale manufacturing experiments, medical studies, and environmental monitoring. Despite their prevalence, the joint modeling of these two features while maintaining computational tractability for large-scale datasets represents a challenging and largely unresolved problem.

Gaussian processes are commonly used for modeling the relationship between inputs and outputs due to their desirable properties. As a flexible and nonparametric approach, Gaussian processes do not require the specification of a fixed functional form and naturally provide uncertainty quantification through their probabilistic framework. However, a major limitation of standard Gaussian process models is their poor scalability to large datasets, as the computational cost grows cubically with the number of observations. To address this challenge, several scalable Gaussian process methods have been developed. The local approximate Gaussian process \citep{gramacy2015local} partitions the input space into smaller subregions and fits independent local models. The nearest neighbor Gaussian process \citep{datta2016hierarchical} and the Vecchia approximation \citep{katzfuss2021general} improve scalability through conditional independence assumptions. Extensions to nonstationary settings have also been proposed, including warping-based methods \citep{schmidt2003bayesian} and deep Gaussian processes \citep{damianou2013deep}, which build hierarchical compositions of Gaussian processes to capture complex input-output relationships. While these methods substantially improve scalability, they assume fully observed responses and do not accommodate censoring or replication structures. 

Recent work has addressed censored Gaussian process models under different inferential frameworks. \citet{groot2012censoredgp} introduced GP regression with a Tobit likelihood and used expectation propagation (EP) to approximate the resulting non-Gaussian posterior. \textcolor{blue}{\citet{gammelli2020estimating} developed a censored-likelihood Gaussian process and estimated it by variational inference, and \citet{gammelli2022generalized} extended this framework to multi-output settings. \citet{karlova2024variational} likewise proposed a variational inference approach for GP models with a Tobit likelihood and derived an analytic evidence lower bound (ELBO) for censored data. In these variational approaches, evaluating the ELBO requires densities of high-dimensional multivariate normal distributions, whose computation becomes expensive or infeasible as the dataset grows; consequently, even where implementations are publicly available, such as the code accompanying \citet{gammelli2020estimating}, they are not directly usable for the large replicated datasets considered here. \citet{hoffmann2025gaussian} developed GP regression methods for value-censored functional and longitudinal data, but their approach relies on standard multivariate normal sampling routines rather than a dedicated large-scale algorithm and is likewise aimed at moderate-sized problems.
These approaches provide principled treatments of censoring, but their reliance on approximate posterior inference or full GP covariance operations limits their direct applicability to large replicated datasets, which is precisely the gap our Vecchia-based construction is designed to fill.} 
\citet{cao2018model} developed an exact likelihood-based framework by integrating over latent truncated Gaussian variables, resulting in multivariate normal probability terms in the likelihood. \citet{chen2022adaptive} built upon this formulation to develop an adaptive experimental design criterion for censored GP models. Although these approaches achieve exact likelihood-based inference under censoring, they require evaluating high-dimensional multivariate normal probabilities and are therefore computationally feasible primarily for small datasets. To improve scalability of censored likelihood evaluation, \citet{cao_katzfuss_2025_vecchia} proposed a Vecchia-based minimax exponential tilting (VMET) method that leverages sparse factorizations to evaluate and sample from high-dimensional truncated multivariate normal distributions with linear complexity. \textcolor{blue}{This substantially improves the scalability of the truncated multivariate normal computations that arise in censored likelihoods. However, their work focuses on efficient truncated multivariate normal probability evaluation and sampling, rather than on the prediction and estimation framework for replicated censored Gaussian process regression considered here; such efficient computations could instead be used as a building block within our framework.}

Other work has addressed replicated observations in GP models. \citet{binois2018practical} investigated repeated measurements and developed methods for modeling inhomogeneous variance across input locations using replicate observations. \citet{patil2025vecchia} proposed a Vecchia-approximated Bayesian heteroskedastic GP model that enables scalable inference for large stochastic simulation experiments with replication. While these methods explicitly account for replicated measurements and incorporate scalable approximations, they assume fully observed responses and do not accommodate censoring.

Motivated by these limitations, we develop a unified framework that simultaneously addresses replication structure, censoring, and computational scalability. Our work targets large, pre-collected datasets in which both exact and censored responses may occur as replicated observations at identical input locations. We model all replicated observations at a given input location jointly in the likelihood, rather than collapsing them by averaging. We integrate this likelihood structure within a Vecchia approximation framework, in which the joint likelihood is decomposed into a product of lower-dimensional conditional distributions. Within each conditional component, censored likelihood contributions require evaluating correlated multivariate normal tail probabilities, which we compute using the separation of variables (SOV) estimator of \citet{genz1992computation} and the minimax exponential tilting (MET) method of \citet{botev2017normal}. We further develop a scalable prediction framework under the Vecchia approximation, in which prediction at a new input location is conditioned on a small set of neighboring observations that may include both exact and censored responses, ensuring we use all available information while maintaining computational efficiency. While \citet{cao2018model} derived predictive distributions under censoring by conditioning on the full joint multivariate normal model, our approach performs prediction within the sparse conditioning structure induced by the Vecchia approximation, making it suitable for large-scale datasets. By jointly addressing replication, censoring, and scalability, the proposed framework provides accurate prediction and reliable uncertainty quantification for complex experimental and manufacturing data.

The remainder of the paper is organized as follows. Section~\ref{sec: method} introduces the proposed modeling framework. Section~\ref{sec: simu} presents numerical studies comparing the proposed method with existing approaches, and Section~\ref{sec: real} illustrates its performance through a real data application. Finally, Section~\ref{sec: disc} concludes the paper and discusses potential directions for future research.

\section{Methods}\label{sec: method}

\subsection{Model}

\subsubsection{Notation}

Let $\mathcal{X} = \{\mathbf{x}_i\}_{i=1}^n$ denote the set of $n$ unique input locations (that is, the distinct experimental settings at which responses are recorded), where each $\mathbf{x}_i \in \mathbb{R}^p$ represents a $p$-dimensional input vector. 
At each input $\mathbf{x}_i$, suppose $r_i \ge 1$ replicate responses are collected, giving a total response of $N = \sum_{i=1}^n r_i$. 
We can denote the response for replicate $j$ at $\mathbf{x}_i$ as $Y_{ij}$, and stack all responses into the vector
$
\mathbf{Y} 
= 
\bigl(Y_{11}, \ldots, Y_{1r_1}, \ldots, Y_{n1}, \ldots, Y_{nr_n}\bigr)^\top 
\in \mathbb{R}^N .
$


For each replicate response at input location $\mathbf{x}_i$, the outcome is either fully observed
(uncensored) or right-censored above a known, input-specific detection limit $C_i$. We define the censoring indicator
$
u_{ij}=\mathbb{I}(Y_{ij}\ge C_i),
$
where the recorded response $Y_{ij}$ is
\[
Y_{ij}=
\begin{cases}
y_{ij}, & Y_{ij}<C_i \quad \text{(uncensored: exact value observed)},\\[4pt]
\text{``}\ge C_i\text{''}, & Y_{ij}\ge C_i \quad \text{(censored: only know }Y_{ij}\ge C_i\text{)}
\end{cases}
\]
For censored observations, we introduce a latent variable $y_{ij}^*$ satisfying $y_{ij}^*\ge C_i$,
representing the unobserved true response underlying the censoring event. Hence, uncensored responses
are observed exactly, while censored responses are only known to exceed $C_i$.

For each $\mathbf{x}_i$, let $m_i = \sum_{j=1}^{r_i} (1-u_{ij})$ denote the number of uncensored replicates, with the remaining $r_i - m_i = \sum_{j=1}^{r_i} u_{ij}$ replicates being censored. 
The replicate-level responses associated with $\mathbf{x}_i$ can then
be partitioned into observed and latent components as
\[
(\mathbf{y}_i, \mathbf{y}_i^*)
=
(y_{i1}, \ldots, y_{i m_i},\, y_{i1}^*, \ldots, y_{i(r_i - m_i)}^*),
\]
where $\mathbf{y}_i$ contains the observed (uncensored) responses and
$\mathbf{y}_i^*$ contains the latent (censored) responses.

Equivalently, we can define the index sets
\[
\mathcal{U} = \{(i,j): u_{ij} = 0\},
\qquad
\mathcal{C} = \{(i,j): u_{ij} = 1\},
\]
and let $\mathbf{y}_{\mathcal{U}}$ and $\mathbf{y}^*_{\mathcal{C}}$ denote the vectors corresponding to the observed uncensored responses and the latent censored responses, respectively.
We can define the stacked vectors
\[
\mathbf{y}_{\mathcal{U}}=\mathbf{y} = \{\mathbf{y}_i\}_{i=1}^n, 
\qquad
\mathbf{y}^*_{\mathcal{C}}=\mathbf{y}^* = \{\mathbf{y}_i^*\}_{i=1}^n,
\]
and the stacked indicator vector $\mathbf{u}=\{\mathbf{u}_i\}_{i=1}^n$ with $\mathbf{u}_i=(u_{i1},\ldots,u_{ir_i})^\top$.
Thus, the dataset can then be expressed as
\[
\mathcal{D} 
= 
\bigl\{\,\mathcal{X},\, \mathbf{y},\, \mathbf{y}^*,\, \mathbf{u}\,\bigr\}
\quad \text{subject to the constraint} \quad
\mathbf{y}^* \ge \mathbf{C},
\]
where \(\mathbf{y}^*\) denotes latent censored responses, $\mathbf{C}$ stacks the corresponding censoring thresholds.

\subsubsection{Gaussian process model}

Given the notations introduced above, we model the response $Y_{ij}$ as
\begin{equation}
    Y_{ij} = W(\mathbf{x}_i) + \epsilon_{ij},
    \label{eq:basicGP}
\end{equation}
where $W(\mathbf{x}_i)$ denotes the underlying latent function evaluated at input $\mathbf{x}_i$. We assume this latent function follows that
$
W \sim \mathcal{GP}\!\bigl(0,\, k(\mathbf{x},\mathbf{x}';\boldsymbol{\theta})\bigr),
$
with zero mean and covariance kernel $k(\cdot,\cdot)$ parameterized by hyperparameters 
$\boldsymbol{\theta}$ (marginal variance $\tau^2$ and length-scale(s) $\ell$). 
The residual term $\epsilon_{ij}$ represents independent measurement noise and is assumed to be,
$
    \epsilon_{ij} \overset{\text{iid}}{\sim} \mathcal{N}(0, \sigma^2).
$

For simplicity, we stack all responses across input locations into the vector 
$\mathbf{Y} = (\mathbf{y}, \mathbf{y}^*)^\top$, which jointly follows
$
(\mathbf{y}, \mathbf{y}^*)^\top \sim 
\mathcal{N}\bigl(\boldsymbol{\mu},\, \boldsymbol{\Sigma}\bigr).
$
Then model~\eqref{eq:basicGP} can be written in vectorized form as
\[
\mathbf{Y} = \mathbf{A}\mathbf{W} + \boldsymbol{\epsilon},
\qquad 
\boldsymbol{\epsilon} \sim \mathcal{N}(\mathbf{0},\, \sigma^2 \mathbf{I}_N),
\]
where $\mathbf{W} = \bigl(W(\mathbf{x}_1), \ldots, W(\mathbf{x}_n)\bigr)^\top \in \mathbb{R}^n$ 
represents the latent GP values at the $n$ unique input locations, and 
$\mathbf{A} \in \mathbb{R}^{N \times n}$ is a block-diagonal matrix  that assigns each replicate observation 
to its corresponding input location. The mean and covariance of $\mathbf{Y}$ are given by $\boldsymbol{\mu} = \mathbf{0}$ and
$
\boldsymbol{\Sigma} 
= 
\mathbf{A} K(\mathcal{X}, \mathcal{X}) \mathbf{A}^\top + \sigma^2 \mathbf{I}_N,
$
where $K(\mathcal{X}, \mathcal{X}) = \bigl\{k(\mathbf{x}_i, \mathbf{x}_h)\bigr\}_{i,h=1}^n$ 
describes the covariance across the unique input locations.

\subsection{Likelihood-based Inference}

\subsubsection{\textcolor{blue}{Vecchia approximation for scalability}}

{\color{blue}
Gaussian process (GP) models are widely used to emulate computer experiments because they yield flexible surrogate surfaces together with predictive uncertainty. For a fixed ordering of the \(n\) distinct input locations, the observed-data likelihood admits the exact chain-rule factorization
\begin{equation}
p(\mathbf{y},\mathbf{u}\mid \boldsymbol{\theta})
=
\prod_{i=1}^{n}
p(\mathbf{y}_i,\mathbf{u}_i
\mid
\mathbf{y}_{<i},\mathbf{u}_{<i};
\boldsymbol{\theta}),
\end{equation}
where
\[
\mathbf{y}_{<i}=(\mathbf{y}_1,\ldots,\mathbf{y}_{i-1}),
\qquad
\mathbf{u}_{<i}=(\mathbf{u}_1,\ldots,\mathbf{u}_{i-1})
\]
Here \(\mathbf{y}_i\) collects the uncensored responses at input location \(i\), and \(\mathbf{u}_i\) the corresponding censoring indicators. The latent censored responses are unobserved, they enter the likelihood only through the censoring events recorded in \(\mathbf{u}_i\).
Although this factorization is exact, the \(i\)th conditional density may depend on all preceding response and censoring blocks. Evaluating it therefore retains the \(\mathcal{O}(N^3)\) cost of the full GP likelihood, where \(N\) is the total number of responses across the \(n\) locations, which is prohibitive for large experiments.

To reduce the computational cost, we apply the Vecchia approximation 
\citep{vecchia1988estimation}, which replaces the full conditioning history with a small set of nearby predecessors. The approximation is constructed at the input-location level, with all replicates observed at the same input location treated as a single block. Locations with at least one uncensored response are ordered before locations with only censored responses, so that directly observed responses are available for conditioning whenever possible.
For each input location \(i\), let
\[
N(i)\subset \{1,\ldots,i-1\},
\qquad
|N(i)|\le k,
\]
denote the set of at most \(k\) nearest preceding locations in the input space. 
We construct \(N(i)\) using only preceding locations that contain at least one 
uncensored response, and write \(\mathbf{y}_{N(i)}\) for the corresponding 
observed responses. The approximate likelihood is then
\begin{equation}
p(\mathbf{y},\mathbf{u}\mid \boldsymbol{\theta})
\approx
\prod_{i=1}^{n}
p\left(
\mathbf{y}_i,\mathbf{u}_i
\mid
\mathbf{y}_{N(i)};
\boldsymbol{\theta}
\right)
\end{equation}
This conditioning structure keeps each local conditional density based on directly observed responses, while censored observations still contribute to the likelihood through their own current-block terms.

Both the nearest-neighbor Gaussian process of \citet{datta2016hierarchical} and the general Vecchia framework of \citet{katzfuss2021general} are based on the Vecchia approximation, and both establish that such factorizations define a valid directed acyclic graph (DAG) approximation corresponding to a proper joint density, rather than a purely heuristic construction. \citet{katzfuss2021general} further distinguish latent and response Vecchia approximations and, for standard Gaussian process regression, recommend the sparse general Vecchia (SGV) construction. In the censored setting we instead use a response-level Vecchia approximation, whose neighbor structure is closer to that of the nearest-neighbor Gaussian process, because each response may be either observed or censored and response-level conditioning is what allows the joint observed-data likelihood to be assembled from local factors. Conditioning a local factor on censored neighbors would render the corresponding conditional distribution non-Gaussian: although its mean and variance remain available in closed form (Proposition~\ref{prop:ynew_cens}), evaluating its density would require an additional integration over the censored neighbors. We therefore condition only on observed responses when fitting the model, while allowing prediction neighborhoods to include both observed and censored responses, as described in the prediction step.
}

\subsubsection{\textcolor{blue}{Likelihood formulation}}

{\color{blue}
We can derive the local conditional contribution
appearing in the Vecchia likelihood. At input location \(i\), let \(r_i\) be the number of replicate responses and let \(m_i\) be the number of uncensored responses. We write \(\mathbf{y}_i\) for the vector of uncensored responses and \(\mathbf{y}_i^*\) for the latent responses corresponding to the censored replicates. Under right censoring at threshold \(C_i\), the censored part is observed through the event
\[
\mathbf{y}_i^* \ge C_i\mathbf{1},
\]
where \(\mathbf{1}\) is a vector of ones of length \(r_i-m_i\).
Given the neighboring responses \(\mathbf{y}_{N(i)}\), the Gaussian process model implies a joint Gaussian distribution for the uncensored responses \(\mathbf{y}_i\) and the latent censored responses \(\mathbf{y}_i^*\). Therefore, the observed-data contribution at location \(i\) can be written as
\begin{equation}
p(\mathbf{y}_i,\mathbf{u}_i \mid \mathbf{y}_{N(i)};\boldsymbol{\theta})
=
p(\mathbf{y}_i \mid \mathbf{y}_{N(i)};\boldsymbol{\theta})
\Pr\left(
\mathbf{y}_i^* \ge C_i\mathbf{1}
\mid
\mathbf{y}_{N(i)},\mathbf{y}_i;\boldsymbol{\theta}
\right)
\label{eq:local_likelihood_decomposition}
\end{equation}
The first factor is the conditional density of the uncensored responses, while the second factor is the conditional probability of the censoring event.

Under the Gaussian process model,
\begin{equation*}
\mathbf{y}_i
\mid
\mathbf{y}_{N(i)},\boldsymbol{\theta}
\sim
\mathcal{N}
\left(
\boldsymbol{\mu}_i^{\mathrm{obs}},
\boldsymbol{\Sigma}_i^{\mathrm{obs}}
\right),
\end{equation*}
where \(\boldsymbol{\mu}_i^{\mathrm{obs}}\) and 
\(\boldsymbol{\Sigma}_i^{\mathrm{obs}}\) are obtained from the standard Gaussian conditioning equations. Similarly, the conditional distribution of the latent censored responses, given the neighboring responses and the uncensored responses at the same location, is
\begin{equation*}
\mathbf{y}_i^*
\mid
\mathbf{y}_{N(i)},\mathbf{y}_i,\boldsymbol{\theta}
\sim
\mathcal{N}
\left(
\boldsymbol{\mu}_i,
\boldsymbol{\Sigma}_i
\right),
\end{equation*}
where \(\boldsymbol{\mu}_i\) and \(\boldsymbol{\Sigma}_i\) denote the corresponding conditional mean vector and covariance matrix. When \(m_i=0\), there are no uncensored responses at location \(i\), so this distribution conditions only on \(\mathbf{y}_{N(i)}\).
The censored component is therefore the multivariate Gaussian tail probability
\begin{equation*}
\Pr\left(
\mathbf{y}_i^* \ge C_i\mathbf{1}
\mid
\mathbf{y}_{N(i)},\mathbf{y}_i;\boldsymbol{\theta}
\right)
\end{equation*}
The resulting censored likelihood contribution requires evaluating a correlated multivariate normal tail probability, which cannot be
factorized into independent terms and is therefore computationally challenging.
}

\subsubsection{\textcolor{blue}{Evaluation of censored probabilities}}

{\color{blue}
To evaluate the censored likelihood contributions described above, we adopt the separation of variables (SOV) method proposed by \citet{genz1992computation}.
This approach exploits the Gaussian conditional distribution of the latent censored responses to efficiently evaluate multivariate normal probabilities under linear inequality constraints.

Let the conditional covariance matrix be decomposed as
\[
\boldsymbol{\Sigma}_i = \mathbf{L}_i\mathbf{L}_i^\top,
\]
where $\mathbf{L}_i \in \mathbb{R}^{(r_i - m_i)\times (r_i - m_i)}$
is the lower-triangular Cholesky factor. With the change of variables
\[
\mathbf{z}_i = \mathbf{L}_i^{-1}(\mathbf{y}_i^{*} - \boldsymbol{\mu}_i),
\]
the original censoring constraint
$\mathbf{y}_i^{*} \ge C_i\mathbf{1}_{r_i - m_i}$
is transformed into integration limits on
$\mathbf{z}_i \sim \mathcal{N}(\mathbf{0}, \mathbf{I}_{r_i - m_i})$
that depend recursively on the previously integrated components.

For \(j=1,\ldots,r_i-m_i\), these transformed bounds are
\begin{align*}
\alpha_{1} 
&\stackrel{\mathrm{def}}{=} \frac{C_i-\mu_{i,1}}{L_{11}}
\;\le\; z_{i,1} \;\le\; \infty, \\[6pt]
\alpha_{2}(z_{i,1})
&\stackrel{\mathrm{def}}{=} \frac{C_i-\mu_{i,2}-L_{21}z_{i,1}}{L_{22}}
\;\le\; z_{i,2} \;\le\; \infty, \\[4pt]
&\vdots \\[4pt]
\alpha_j(z_{i,1},\ldots,z_{i,j-1})
&\stackrel{\mathrm{def}}{=}
\frac{
C_i-\mu_{i,j}-\sum_{k=1}^{j-1}L_{jk}z_{i,k}
}{
L_{jj}
}
\le z_{i,j}<\infty 
\end{align*}
Thus, the lower bound for \(z_{i,j}\) depends on the previously integrated components \(z_{i,1},\ldots,z_{i,j-1}\), reflecting the sequential structure induced by the lower-triangular matrix \(\mathbf{L}_i\).
Under this transformation, the multivariate normal probability can be written as an iterated integral of univariate standard normal densities:
\begin{equation}
\begin{aligned}
&\Pr\left(
\mathbf{y}_i^*\ge C_i\mathbf{1}
\mid
\mathbf{y}_{N(i)},\mathbf{y}_i
\right)  \\
&\quad =
\int_{\alpha_1}^{\infty}
\phi(z_{i,1})
\int_{\alpha_2(z_{i,1})}^{\infty}
\phi(z_{i,2})
\cdots
\int_{\alpha_{r_i-m_i}(z_{i,1},\ldots,z_{i,r_i-m_i-1})}^{\infty}
\phi(z_{i,r_i-m_i})\,d\mathbf{z}_i ,
\end{aligned}
\label{eq:sov_integral}
\end{equation}
where \(\phi(\cdot)\) denotes the standard normal density.
When \(r_i-m_i\) is large, direct numerical integration of
\eqref{eq:sov_integral} can be computationally expensive. We therefore use a Monte Carlo approximation based on the SOV representation. The method retains the sequential one-dimensional structure of the integral, but estimates the probability by Monte Carlo sampling. At each step, \(z_{i,j}\) is sampled from a standard normal distribution truncated below at
\(\alpha_j(z_{i,1},\ldots,z_{i,j-1})\). For a simulated path
\(\mathbf{z}_i=(z_{i,1},\ldots,z_{i,r_i-m_i})^\top\), the corresponding Monte Carlo SOV weight is
\[
w =
\prod_{j=1}^{r_i-m_i}
\left[
1 -
\Phi\!\left\{
\alpha_j(z_{i,1},\ldots,z_{i,j-1})
\right\}
\right],
\]
where \(\Phi(\cdot)\) denotes the standard normal distribution function. The censored probability is then estimated by averaging these weights over Monte Carlo samples. 
The inverse-transform derivation of the truncated-normal sampling step is given in Appendix~\ref{app:sov_inverse}.
}

\begin{algorithm}[H]
\caption{\small MC–SOV}
\footnotesize
\setlength{\abovedisplayskip}{0pt}
\setlength{\belowdisplayskip}{0pt}
\setlength{\abovedisplayshortskip}{0pt}
\setlength{\belowdisplayshortskip}{0pt}
\setlength{\lineskiplimit}{0pt}
\setlength{\lineskip}{0pt}
\setlength{\jot}{0pt}
\begin{algorithmic}
\Require $\mathbf{L}_i$ (Cholesky of $\boldsymbol{\Sigma}_i$), $\boldsymbol{\mu}_i$, threshold $C_i$, and $M$ Monte Carlo samples

\For{$m = 1, \ldots, M$}
  \For{$j = 1, \ldots, d_i$}, where $d_i = r_i - m_i$
  \State Compute the conditional lower bound:
  \[
  \alpha_{j} = 
  \frac{C_i - \mu_{i,j} - \sum_{k<j} L_{jk} z_{i,k}^{(m)}}{L_{jj}}
  \]
  \State Sample from the truncated normal:
  \[
  z_{i,j}^{(m)} = 
  \Phi^{-1}\!\left[\Phi(\alpha_{j}) + 
  U_{j}(1 - \Phi(\alpha_{j}))\right],
  \qquad U_{j} \sim \mathrm{Unif}(0,1)
  \]
  \State Store the conditional probability weight:
  \[
  w_{j}^{(m)} = 1 - \Phi(\alpha_{j})
  \]
  \EndFor
  \State Compute joint weight:
  \[
  w^{(m)} = \prod_{j=1}^{r_i-m_i} w_{j}^{(m)}
  \]
\EndFor
\State \textbf{Return:}
\[
\displaystyle \widehat{p} =\frac{1}{M}\sum_{m=1}^{M} w^{(m)}
\]
\end{algorithmic}
\end{algorithm}

{\color{blue}
For rare tail probabilities, the Monte Carlo SOV estimator may become inefficient because the product of sequential conditional tail probabilities can be very small, leading to unstable
Monte Carlo weights. To improve stability, we use the minimax exponential tilting (MET) method of \citet{botev2017normal}. 
MET is an importance sampling method for multivariate normal probabilities under linear inequality constraints, and is especially useful when the truncation region lies in the far tail.

Let \(d_i=r_i-m_i\), and define the truncation region induced by the SOV transformation as
\[
\mathcal{R}_i
=
\left\{
\mathbf{z}_i:
z_{i,j}\ge
\alpha_j(z_{i,1},\ldots,z_{i,j-1}),
\quad j=1,\ldots,d_i
\right\}
\]
Then the censored likelihood contribution can be written as
\[
p_i
=
\int_{\mathcal{R}_i}
\phi(\mathbf{z}_i;\mathbf{0},\mathbf{I}_{d_i})\,d\mathbf{z}_i 
\]

MET improves the standard SOV proposal by exponentially tilting each sequential truncated normal component. Specifically, for a tilting parameter \(\boldsymbol{\lambda}_i\in\mathbb{R}^{d_i}\), the tilted conditional proposal for the \(j\)th component is
\[
g_j
\left(
z_{i,j}\mid z_{i,1},\ldots,z_{i,j-1};
\boldsymbol{\lambda}_i
\right)
=
\frac{
\phi(z_{i,j};\lambda_{i,j},1)
I\left\{
z_{i,j}\ge
\alpha_j(z_{i,1},\ldots,z_{i,j-1})
\right\}
}{
1-\Phi\left[
\alpha_j(z_{i,1},\ldots,z_{i,j-1})-\lambda_{i,j}
\right]
}
\]
The resulting joint tilted SOV proposal is
\[
g_{\boldsymbol{\lambda}_i}(\mathbf{z}_i)
=
\prod_{j=1}^{d_i}
g_j
\left(
z_{i,j}\mid z_{i,1},\ldots,z_{i,j-1};
\boldsymbol{\lambda}_i
\right),
\qquad
\mathbf{z}_i\in\mathcal{R}_i 
\]
Under this proposal, the importance weight is
\[
w(\mathbf{z}_i;\boldsymbol{\lambda}_i)
=
\frac{
\phi(\mathbf{z}_i;\mathbf{0},\mathbf{I}_{d_i})
}{
g_{\boldsymbol{\lambda}_i}(\mathbf{z}_i)
}
=
\exp\left\{
\psi_i(\mathbf{z}_i,\boldsymbol{\lambda}_i)
\right\},
\]
where
\[
\psi_i(\mathbf{z}_i,\boldsymbol{\lambda}_i)
=
-\mathbf{z}_i^\top\boldsymbol{\lambda}_i
+
\frac{1}{2}\|\boldsymbol{\lambda}_i\|^2
+
\sum_{j=1}^{d_i}
\log
\left[
1-\Phi
\left\{
\alpha_j(z_{i,1},\ldots,z_{i,j-1})-\lambda_{i,j}
\right\}
\right]
\]
The optimal tilting parameter \(\boldsymbol{\lambda}_i^*\) is chosen using the minimax criterion of \citet{botev2017normal}, which controls the worst-case likelihood ratio over
\(\mathcal{R}_i\). Details on the minimax optimization used to obtain \(\boldsymbol{\lambda}_i^*\) are provided in Appendix~\ref{app:met_details}.
After \(\boldsymbol{\lambda}_i^*\) is obtained, samples are generated from the tilted proposal 
\(g_{\boldsymbol{\lambda}_i^*}\). For a draw
\(\mathbf{z}_i^{(m)}\sim g_{\boldsymbol{\lambda}_i^*}\), the corresponding importance weight is
\[
w(\mathbf{z}_i^{(m)})
=
\frac{
\phi(\mathbf{z}_i^{(m)};\mathbf{0},\mathbf{I}_{d_i})
}{
g_{\boldsymbol{\lambda}_i^*}(\mathbf{z}_i^{(m)})
}
=
\exp
\left\{
\psi_i(\mathbf{z}_i^{(m)},\boldsymbol{\lambda}_i^*)
\right\}
\]
Therefore, given samples
\(\mathbf{z}_i^{(m)}\sim g_{\boldsymbol{\lambda}_i^*}\), \(m=1,\ldots,M\), the MET estimator of the censored probability is
\[
\widehat p_i
=
\frac{1}{M}
\sum_{m=1}^{M}
w(\mathbf{z}_i^{(m)})
=
\frac{1}{M}
\sum_{m=1}^{M}
\exp
\left\{
\psi_i(\mathbf{z}_i^{(m)},\boldsymbol{\lambda}_i^*)
\right\}
\]
}

\begin{algorithm}[H]
\caption{MET}
\footnotesize
\setlength{\abovedisplayskip}{2pt}
\setlength{\belowdisplayskip}{2pt}
\setlength{\jot}{2pt}
\begin{algorithmic}

\Require Cholesky factor \(\mathbf{L}_i\) of 
\(\boldsymbol{\Sigma}_i=\mathbf{L}_i\mathbf{L}_i^\top\); mean 
\(\boldsymbol{\mu}_i\); threshold \(C_i\); number of samples \(M\).
\State Set \(d_i=r_i-m_i\), and define the truncation region
\[
\mathcal{R}_i
=
\left\{
\mathbf{z}_i:
z_{i,j}\ge
\alpha_j(z_{i,1},\ldots,z_{i,j-1}),
\quad j=1,\ldots,d_i
\right\},
\]
where
\[
\alpha_j(z_{i,1},\ldots,z_{i,j-1})
=
\frac{
C_i-\mu_{i,j}-\sum_{k=1}^{j-1}L_{i,jk}z_{i,k}
}{
L_{i,jj}
}
\]

\State \textbf{Step 1: Define the minimax objective function}
\[
\psi_i(\mathbf{z}_i,\boldsymbol{\lambda}_i)
=
-\mathbf{z}_i^{\!\top}\boldsymbol{\lambda}_i
+\tfrac12\|\boldsymbol{\lambda}_i\|^2
+\sum_{j=1}^{d_i}
\ln\!\Big[
1-\Phi\!\big(
\alpha_j(z_{i,1},\ldots,z_{i,j-1})-\lambda_{i,j}
\big)
\Big]
\]

\State \textbf{Step 2: Find the saddle point \((\mathbf{z}_i^*,\boldsymbol{\lambda}_i^*)\)}
Solve
\[
(\mathbf{z}_i^*,\boldsymbol{\lambda}_i^*)
=
\arg\max_{\mathbf{z}_i,\boldsymbol{\lambda}_i}
\psi_i(\mathbf{z}_i,\boldsymbol{\lambda}_i),
\qquad
\text{subject to }
\frac{\partial \psi_i}{\partial \boldsymbol{\lambda}_i}
=
\mathbf{0},
\quad
\mathbf{z}_i\in\mathcal{R}_i,
\]
using:
\begin{enumerate}
  \item Powell's dogleg trust-region method; \emph{if} it fails or the resulting candidate is infeasible,
  \item switch to a constrained augmented-Lagrangian solver.
\end{enumerate}

\State \textbf{Step 3: Importance sampling under the tilt}
\For{\(m=1,\ldots,M\)}
  \State Draw
  \[
  \mathbf{z}_i^{(m)}
  \sim
  g_{\boldsymbol{\lambda}_i^*}(\mathbf{z}_i)
  \]
  \State Compute the weight
  \[
  w^{(m)}
  =
  \exp\!\left(
  -{\boldsymbol{\lambda}_i^*}^{\!\top}\mathbf{z}_i^{(m)}
  +
  \tfrac12\|\boldsymbol{\lambda}_i^*\|^2
  +
  \sum_{j=1}^{d_i}
  \ln
  \left[
  1-\Phi
  \left\{
  \alpha_j(z_{i,1}^{(m)},\ldots,z_{i,j-1}^{(m)})
  -
  \lambda_{i,j}^*
  \right\}
  \right]
  \right)
  \]
\EndFor

\State \textbf{Return:}
\[
\widehat{p}_i
=
\frac{1}{M}
\sum_{m=1}^{M}
w^{(m)}
\]
\end{algorithmic}
\end{algorithm}

\subsubsection{\textcolor{blue}{Parameter estimation}}

{\color{blue}
We estimate the model parameters by maximizing the Vecchia approximate observed-data log-likelihood. Combining the local contributions over all input locations derived above, we have
\begin{equation}
\log \mathcal{L}(\boldsymbol{\theta}\mid \mathbf{y},\mathbf{u})
\approx
\sum_{i=1}^{n}
\log \phi_{m_i}\left(y_i;\boldsymbol{\mu}^{obs}_i,\boldsymbol{\Sigma}^{obs}_i\right)
+
\sum_{i=1}^{n}
\log
\Pr\left(
\mathbf{y}_i^* \ge C_i\mathbf{1}
\mid
\mathbf{y}_{N(i)},\mathbf{y}_i;\boldsymbol{\theta}
\right)
\label{eq:log_likelihood_vecchia}
\end{equation}
where \(\phi_{m_i}\) is the \(m_i\)-dimensional Gaussian density. The first term is the contribution from the uncensored responses, and the second term is the contribution from the censored responses through the corresponding Gaussian tail probability. When \(m_i=0\), the first term contributes zero to the log-likelihood; when \(r_i-m_i=0\), the second term contributes zero.
The unknown parameters in this likelihood are collected in
\[
\boldsymbol{\theta}=(\ell,\tau^2,\sigma^2,\boldsymbol{\beta}),
\]
where \(\ell\) is the lengthscale parameter, \(\tau^2\) is the variance, \(\sigma^2\) is the nugget, and \(\boldsymbol{\beta}\) contains the mean-function coefficients. For a zero-mean Gaussian process, \(\boldsymbol{\beta}\) is omitted. When the data exhibit a clear linear trend, we define the mean as \(\boldsymbol{\mu} = (1, \mathbf{x}^\top)\boldsymbol{\beta}\), where \(\boldsymbol{\beta}\) includes an intercept and linear coefficients corresponding to the input covariates.
In practice, we estimate \(\boldsymbol{\theta}\) by minimizing the negative approximate log-likelihood in \eqref{eq:log_likelihood_vecchia} using the box-constrained \texttt{L-BFGS-B} algorithm implemented in the \texttt{optim} function in \texttt{R}.
}




\subsubsection{\textcolor{blue}{Prediction}}

{\color{blue}
Given the estimated parameters $\boldsymbol{\hat{\theta}}$, our goal is to predict the response
\(y_{\new}\) at an unobserved input location \(\mathbf{x}_*\). Under the Vecchia framework, prediction is based on a local conditioning set rather than the full training data. Let \(N(*)\) denote the selected neighborhood for \(\mathbf{x}_*\), and partition the neighboring responses as
\[
\mathbf y_{N(*)}
=
\left(\mathbf y_o^\top,\mathbf y_c^\top\right),
\]
where \(\mathbf y_o\) contains the uncensored neighboring responses and \(\mathbf y_c\) denotes the latent responses corresponding to censored neighboring observations. The observed information from the censored neighbors is the event
\[
\mathbf y_c\ge\mathbf C,
\]
where \(\mathbf C\) is the vector of corresponding censoring thresholds.
The neighborhood \(N(*)\) is first chosen as the \(k\) nearest training locations to \(\mathbf{x}_*\) in the input space. To avoid conditioning on a neighborhood dominated by censored observations, we impose an upper bound on the number of censored neighbors. If the initially selected neighborhood contains more censored neighbors than this upper bound, the farthest censored neighbors are removed first and replaced by the nearest available uncensored observations outside the initial neighborhood, using the same input-space distance criterion. This procedure keeps the total neighborhood size equal to \(k\), while ensuring that sufficient directly observed responses are available for conditioning. In our implementation, the number of censored neighbors is limited to at most five.


{\color{blue}
When all selected neighbors are uncensored, the predictive distribution of \(y_{\new}\) is the usual Gaussian conditional distribution under the Vecchia approximation. When censored neighbors are present, the censoring information is incorporated through the conditional distribution of the latent responses \(\mathbf y_c\) given \(\mathbf y_o\), truncated to the region \(\mathbf y_c\ge\mathbf C\). 
The resulting predictive distribution of \(y_{\new}\) is no longer Gaussian, but its first two moments are available in closed form in terms of the mean and covariance of this truncated multivariate normal distribution. The following proposition gives these predictive moments. Related moment-based expressions for censored Gaussian process models were developed by \citet{cao2018model} and subsequently used by \citet{chen2022adaptive} for adaptive design under censoring. Here, these expressions are applied within the Vecchia framework, with conditioning restricted to a local neighborhood.
}
\begin{proposition}[Predictive moments under censoring]
\label{prop:ynew_cens}
Given the local neighborhood \(N(*)\), the moment-adjusted predictive mean and variance of \(y_{\new}\) conditional on \(\mathbf y_o\) and the censoring event \(\mathbf y_c\ge\mathbf C\) are
\begin{align}
\boldsymbol\mu_{\new}^{\ast}
&= \boldsymbol\mu_{\new\mid o} 
  + \boldsymbol{\Sigma}_{\new,c\mid o}\boldsymbol{\Sigma}_{cc\mid o}^{-1}
    \bigl(\mathbf{m}_{c} - \boldsymbol{\mu}_{c\mid o}\bigr),
\label{eq:mu_star_formal}\\[4pt]
\boldsymbol\Sigma_{\new}^{\ast}
&= \boldsymbol\Sigma_{\new\mid o}
  - \boldsymbol{\Sigma}_{\new,c\mid o}\boldsymbol{\Sigma}_{cc\mid o}^{-1}
    \boldsymbol{\Sigma}_{c,\new\mid o}
  + \boldsymbol{\Sigma}_{\new,c\mid o}\boldsymbol{\Sigma}_{cc\mid o}^{-1}
    \mathbf{V}_{c}\boldsymbol{\Sigma}_{cc\mid o}^{-1}
    \boldsymbol{\Sigma}_{c,\new\mid o}
\label{eq:Sig_star_formal}
\end{align}
\end{proposition}

Here, \(\mathbf{m}_c\) and \(\mathbf{V}_c\) denote the conditional mean and covariance of the censored latent block \(\mathbf y_c\) given \(\mathbf y_o\) and \(\mathbf y_c\ge\mathbf C\). 
{\color{blue}
Although Proposition~\ref{prop:ynew_cens} provides the predictive mean and variance, the predictive distribution under censoring is generally non-Gaussian and belongs to the closed skew-normal family \citep{DominguezMolinaEtAl2003,GonzalezFariasEtAl2004}. Its quantiles therefore cannot be obtained from the mean and variance alone. We construct prediction intervals using the following two-stage composition sampling procedure.

\begin{algorithm}[H]
\color{blue}
\footnotesize
\setlength{\abovedisplayskip}{0pt}
\setlength{\belowdisplayskip}{0pt}
\setlength{\abovedisplayshortskip}{0pt}
\setlength{\belowdisplayshortskip}{0pt}
\setlength{\lineskiplimit}{0pt}
\setlength{\lineskip}{0pt}
\setlength{\jot}{0pt}

\caption{\textcolor{blue}{Composition sampling for prediction with censored neighbors}}
\label{alg:prediction_sampling}
\begin{algorithmic}
\Require Number of draws \(B\), censoring thresholds \(\mathbf C\), and the conditional
moments \(\boldsymbol\mu_{c\mid o},\boldsymbol\Sigma_{cc\mid o},\boldsymbol\mu_{\new\mid o},\boldsymbol\Sigma_{\new,c\mid o},\boldsymbol\Sigma_{\new\mid o}\)

\For{\(b=1,\ldots,B\)}

    \Statex \textbf{Step 1: Sample the latent censored responses.}

    \Statex Draw
    \[
    \mathbf y_c^{(b)}
    \sim
    \mathcal N\!\left(
    \boldsymbol\mu_{c\mid o},
    \boldsymbol\Sigma_{cc\mid o}
    \right)
    \quad\text{truncated to }
    \{\mathbf y_c \ge \mathbf C\}
    \]

    \Statex \textbf{Step 2: Sample the new response.}

    \Statex Conditional on \(\mathbf y_c^{(b)}\), draw
    \[
    y_{\new}^{(b)}
    \sim
    \mathcal N\!\left(
    \boldsymbol\mu_{\new\mid o}
    +
    \boldsymbol\Sigma_{\new,c\mid o}
    \boldsymbol\Sigma_{cc\mid o}^{-1}
    \left(
    \mathbf y_c^{(b)}
    -
    \boldsymbol\mu_{c\mid o}
    \right),
    \;
    \boldsymbol\Sigma_{\new\mid o}
    -
    \boldsymbol\Sigma_{\new,c\mid o}
    \boldsymbol\Sigma_{cc\mid o}^{-1}
    \boldsymbol\Sigma_{c,\new\mid o}
    \right)
    \]

\EndFor
\Statex \textbf{Step 3: Construct the prediction interval.}
\Statex For a \(100(1-\alpha)\%\) prediction interval, compute the
empirical \(\alpha/2\) and \(1-\alpha/2\) quantiles of
\[
y_{\new}^{(1)},\ldots,y_{\new}^{(B)}
\]
\end{algorithmic}
\end{algorithm}

The derivation of Proposition~\ref{prop:ynew_cens} and the computation of the truncated moments are provided in Appendices~\ref{app:prediction_derivation} and \ref{app:prediction_moment}, respectively. Appendix~\ref{app:prediction_sampling} derives the sampler in Algorithm~\ref{alg:prediction_sampling}.
}
}

\subsection{\textcolor{blue}{Computational complexity}}

The computational cost of the proposed approach can be decomposed into two parts:
(i) likelihood evaluation for parameter estimation and
(ii) prediction at new input locations. Both parts exploit the local conditioning structure of the Vecchia approximation, together with efficient evaluation of censored probabilities.

\paragraph{\textcolor{blue}{Estimation}}

{\color{blue}
For the uncensored component of the likelihood, the Vecchia approximation replaces the full joint density with a product of local conditional densities, each involving at most \(k\) nearest neighbors. Evaluating one local Gaussian conditional density requires a Cholesky factorization of a covariance matrix of size at most \(k\), with cost \(\mathcal{O}(k^3)\). Therefore, over \(n\) unique input locations, the uncensored likelihood contribution has cost $\mathcal{O}(n k^3)$.

For the censored component, let \(d_i\) denote the number of censored replicates at input location \(i\). The corresponding likelihood contribution requires evaluating a \(d_i\)-dimensional truncated multivariate normal probability. When MC--SOV is used, this calculation involves a Cholesky factorization of the local conditional covariance matrix, with cost \(\mathcal{O}(d_i^3)\), followed by Monte Carlo evaluation with cost \(\mathcal{O}(M d_i^2)\), where \(M\) is the number of Monte Carlo samples. Thus, the MC--SOV cost for the censored contribution at location \(i\) is $\mathcal{O}(d_i^3+M d_i^2)$.
When MET is used, the censored probability is evaluated using an exponentially tilted proposal. The sampling step has the same order as the SOV-based sequential evaluation, but MET also requires an additional numerical optimization step to obtain the tilting parameter. In the original MET formulation, this optimization is iterative and has cubic per-iteration cost in the dimension of the truncated multivariate normal probability. In our setting, this optimization is applied only to the local \(d_i\)-dimensional censored block. Its cost is therefore kept separate from the Monte Carlo sampling cost, since it depends on the censored block size and on the convergence behavior of the optimizer.

Consequently, when MC--SOV is used to evaluate the censored probabilities, the total computational cost of the approximate likelihood calculation is
\[
\mathcal{O}\!\left(
n k^3 + \sum_{i=1}^n \left(d_i^3 + M d_i^2\right)
\right)
\]
When MET is used, this cost is augmented by the additional local optimization required to obtain the tilting parameters for the censored blocks.

In both cases, the Monte Carlo sample size \(M\) is a fixed, prespecified computational setting rather than an adaptively chosen quantity, so that each censored probability is estimated with error decreasing at the usual Monte Carlo rate; this follows the practice of \citet{cao_katzfuss_2025_vecchia}, who likewise fix the Monte Carlo sample size in their linear-cost tilting-based estimator. The MET tilting parameter is obtained from the minimax problem of \citet{botev2017normal}, whose tilted proposal has the vanishing-relative-error property and is therefore numerically stable even far in the tail; because we apply this optimization only to the low-dimensional local censored blocks, both its cost and the number of optimizer iterations remain small in our setting.
}

\paragraph{Prediction}
Prediction at a new input location \(\mathbf{x}_*\) is performed under the same local Vecchia conditioning structure. When all neighboring responses are uncensored, prediction requires a Gaussian conditional calculation based on a local covariance matrix of size at most \(k\). Therefore, the per-prediction cost is $\mathcal{O}(k^3)$.

{\color{black}
When censored neighbors are included, let \(d_{c,i}\) denote the number of censored neighbors used for the \(i\)th prediction. Computing the closed-form predictive mean and variance requires the truncated moments of the censored-neighbor block. This involves a Cholesky factorization of the \(d_{c,i}\times d_{c,i}\) conditional covariance matrix, with cost \(\mathcal{O}(d_{c,i}^3)\), and Monte Carlo evaluation of the truncated
moments, with cost \(\mathcal{O}(M d_{c,i}^2)\). Therefore, the cost of computing the predictive mean and variance is
\[
\mathcal{O}\!\left(
k^3+d_{c,i}^3+M d_{c,i}^2
\right)
\]

Constructing prediction intervals requires an additional composition sampling step. Let \(B\) denote the number of predictive draws. For each draw, a \(d_{c,i}\)-dimensional truncated normal vector is generated using the exact minimax-tilting method of \citet{botev2017normal}, followed by a
conditional Gaussian draw of \(y_{\new}\). The latter calculation has cost \(\mathcal{O}(B d_{c,i})\), whereas the cost of generating the truncated normal draws depends on the implementation and acceptance behavior of the minimax-tilting sampler. Because \(d_{c,i}\) is restricted to at most five
in our implementation, this additional sampling step remains
low-dimensional.

Overall, the proposed method avoids direct computation with the full \(N\times N\) covariance matrix by replacing the full GP calculation with local Gaussian conditioning, low-dimensional truncated-moment calculations, and low-dimensional composition sampling.
}

\section{Simulation Study} \label{sec: simu}

\subsection{Illustrative one dimensional input setup}

To provide visual intuition for how different censoring strategies affect GP fitting, we first examine a simple one-dimensional example that illustrates the key advantages of the proposed framework. 
The dataset consists of 800 unique input locations with input values $x_i$ evenly spaced between 0 and 1. 
Each input location is randomly replicated between one and three times to account for repeated measurements. 
The true underlying function is defined as
\[
f(x) = \sin\{2\times(7x - 2)\} + 2\exp\{-15^2(x - 0.5)^2\} + \sin\{2 \times(2x - 0.5)\}
\]
and the noisy observations are generated according to 
\[
y_{ij} = f(x_i) + \varepsilon_{ij}, \qquad \varepsilon_{ij} \stackrel{\text{iid}}{\sim} \mathcal{N}(0, 0.2^2),
\]
where $\varepsilon_{ij}$ denotes independent noise following a normal distribution with mean 0 and standard deviation 0.2.

\begin{figure}[H]
\centering
\begin{subfigure}{0.45\textwidth}
    \includegraphics[width=\linewidth]{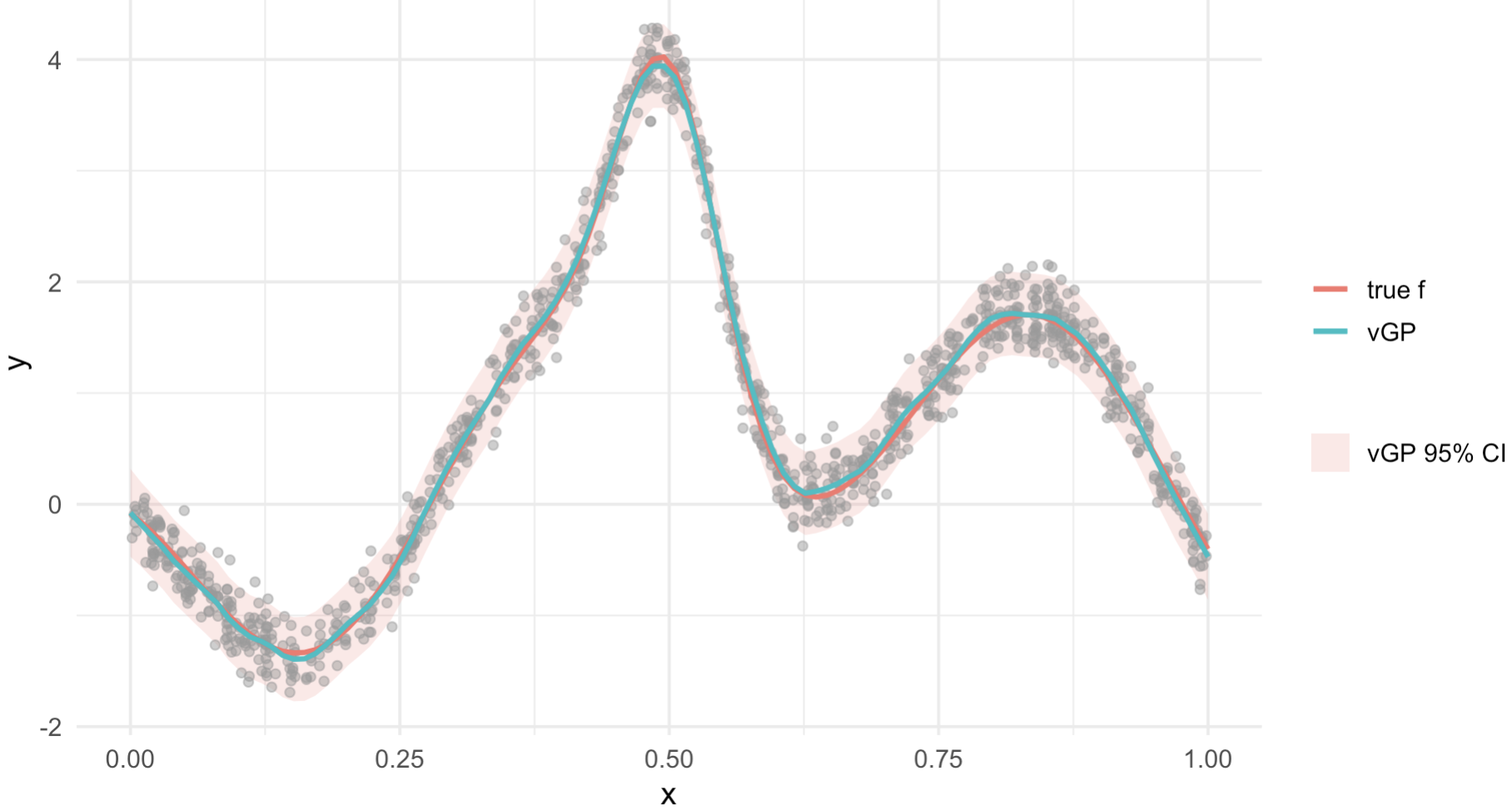}
    \caption{Non-censored GP fit}
    \label{fig:gp1}
\end{subfigure}
\begin{subfigure}{0.45\textwidth}
    \includegraphics[width=\linewidth]{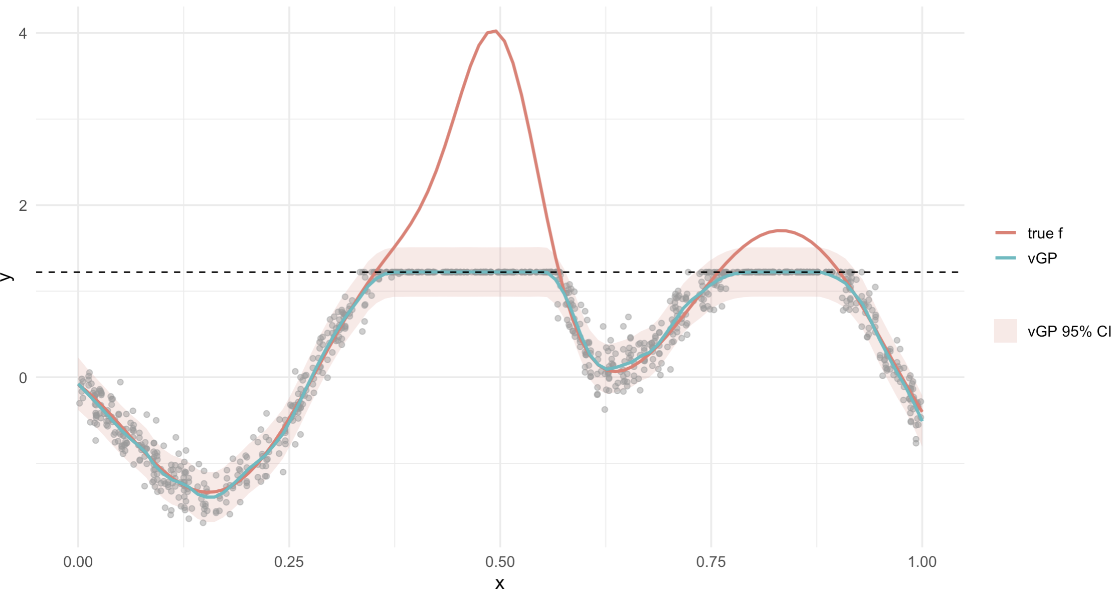}
    \caption{Fix Censored at $C=1.221$ (flat region)}
    \label{fig:gp2}
\end{subfigure}

\vspace{1em} 

\begin{subfigure}{0.45\textwidth}
    \includegraphics[width=\linewidth]{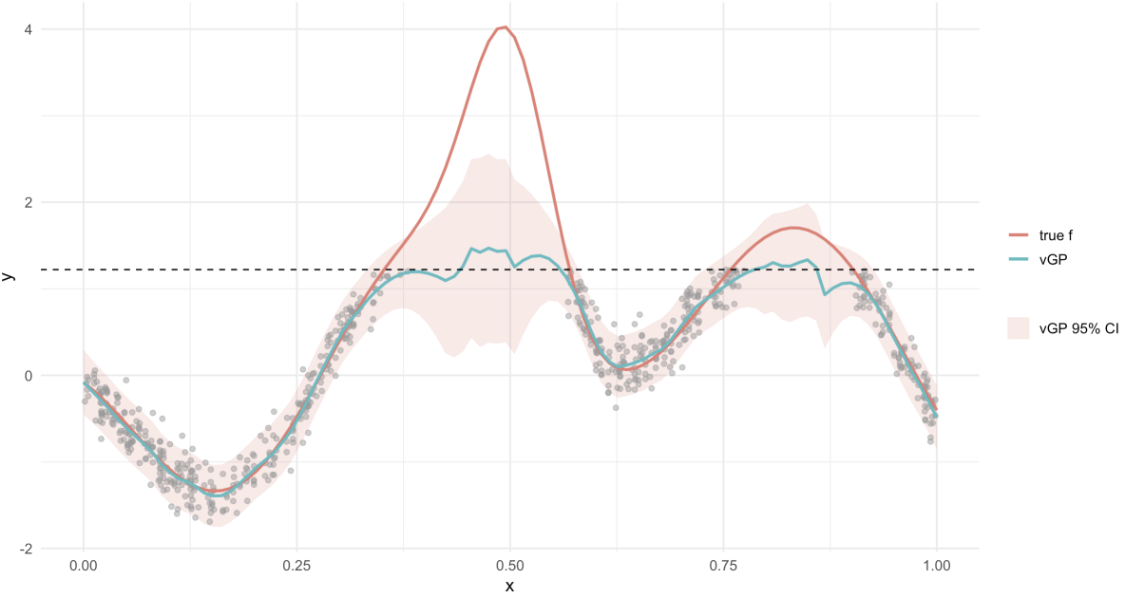}
    \caption{Remove Censored}
    \label{fig:gp3}
\end{subfigure}
\begin{subfigure}{0.49\textwidth}
    \includegraphics[width=\linewidth,height=0.206\textheight]{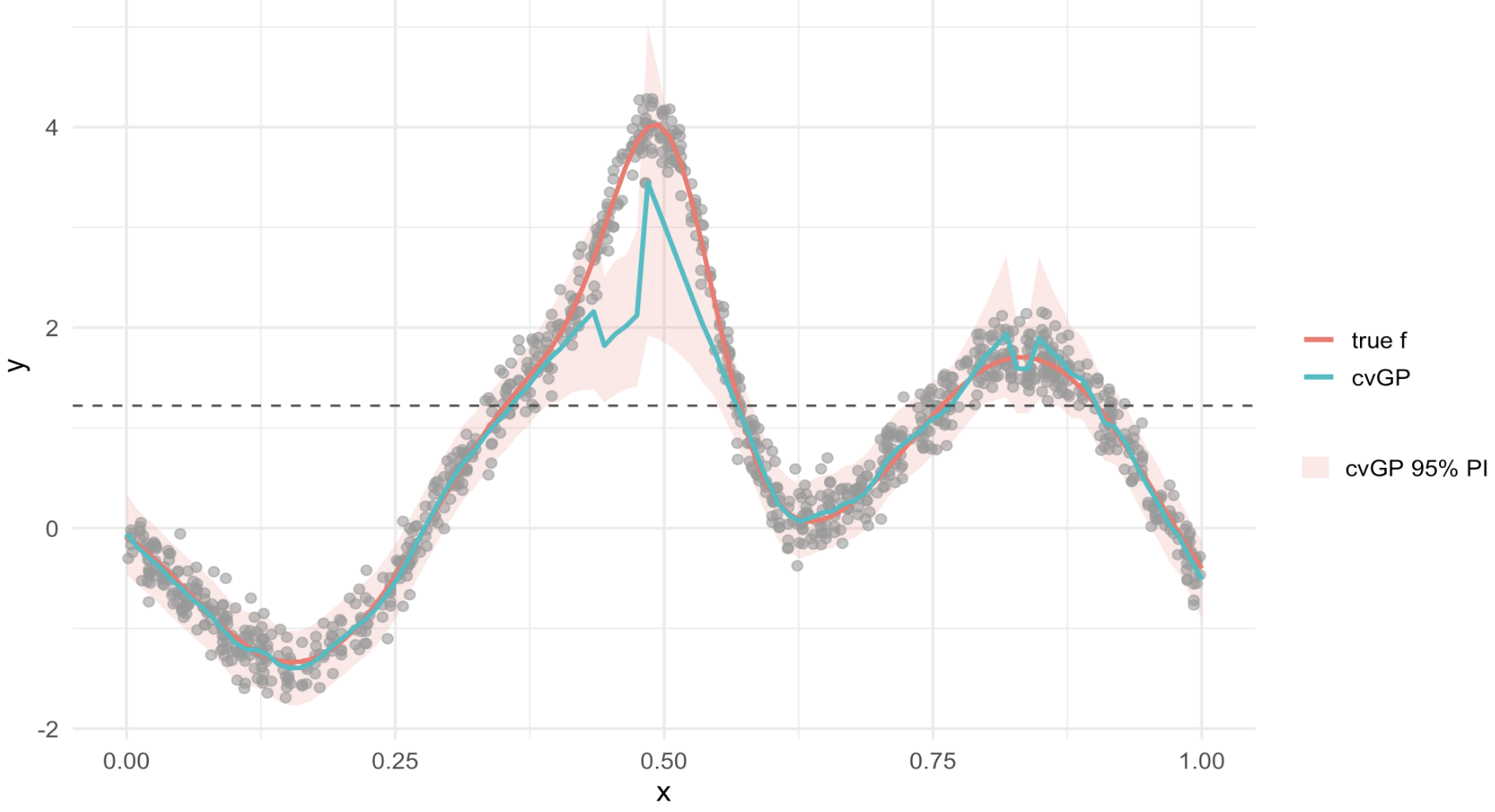}
    \caption{Proposed censored-likelihood GP fit}
    \label{fig:gp4}
\end{subfigure}
\captionsetup{font=footnotesize}
\caption{
Fitted GP outcomes under different censoring scenarios. 
Each panel shows the true function (red), fitted function (blue),
95\% predictive interval (shaded), noisy observations (grey points),
and the censoring threshold at $C = 1.221$ (black dashed line).
}
\label{fig:gp-comparison}
\end{figure}

As shown in Figure~\ref{fig:gp-comparison}, panel~(a) presents the fitted GP curve under the setting without censoring, where the model accounts only for the observed likelihood as defined in Equation~(2). 
In panel~(b), we set a censoring threshold of $C = 1.221$ (dashed line), and all censored observations are fixed at $C$ before fitting the  GP model based solely on the observed likelihood. 
Panel~(c) shows the result when all censored observations are removed from the dataset before fitting the model. 
Finally, panel~(d) displays the proposed GP model that explicitly incorporates the censored likelihood into the inference.  

Comparing panels~(b)–(d), the results differ depending on how the censored part of the data is handled.
When censored observations are fixed at $C$ (panel~b), the model misleadingly treats them as exact measurements, resulting in an unrealistically flat posterior mean and underestimated uncertainty. 
When censored data are removed (panel~c), the model fails to learn from the missing region and exhibits overly wide uncertainty. 
In contrast, our proposed method (panel~d) appropriately captures the underlying trend and provides coherent uncertainty quantification across both observed and censored regions.
\textcolor{blue}{Using this same one-dimensional example, a sensitivity analysis for the maximum number of censored nearest neighbors used in prediction is reported in Appendix~\ref{app:neighbor_sensitivity}.}

\subsection{Simulation Setup}
We carry out a simulation study using the well-known \textit{Borehole function}, which models the water flow rate (in m$^3$/yr) through a borehole based on eight physical input variables. The range, distribution, and description of each input are summarized in Table~\ref{tab:borehole-inputs}.

\begin{table}[h!]
\centering
\footnotesize
\begin{tabular}{llll}
\toprule
Variable & Range & Distribution & Description (units) \\
\midrule
$r_w$ & $[0.05, 0.15]$ & $\mathcal{N}(0.10, 0.0161812^2)$ & radius of borehole (m) \\
$r$   & $[100, 50000]$ & $\mathrm{Lognormal}(7.71, 1.0056^2)$ & radius of influence (m) \\
$T_u$ & $[63070, 115600]$ & $\mathrm{Unif}[63070,\,115600]$ & transmissivity of upper aquifer (m$^2$/yr) \\
$H_u$ & $[990,1110]$ & $\mathrm{Unif}[990,1110]$ & potentiometric head of upper aquifer (m) \\
$T_l$ & $[63.1, 116]$ & $\mathrm{Unif}[63.1, 116]$ & transmissivity of lower aquifer (m$^2$/yr) \\
$H_l$ & $[700, 820]$ & $\mathrm{Unif}[700, 820]$ & potentiometric head of lower aquifer (m) \\
$L$   & $[1120, 1680]$ & $\mathrm{Unif}[1120, 1680]$ & length of borehole (m) \\
$K_w$ & $[9855, 12045]$ & $\mathrm{Unif}[9855, 12045]$ & hydraulic conductivity of borehole (m/yr) \\
\bottomrule
\end{tabular}
\captionsetup{font=footnotesize}
\caption{Input domains and distributions}
\label{tab:borehole-inputs}
\end{table}

The borehole response function is defined by the following equation:
\begin{equation*}
f(\mathbf{x})
= \frac{2\pi\, T_u \,(H_u - H_l)}%
       {\ln\!\bigl(r/r_w\bigr)\,\left(
           1
           + \dfrac{2 L T_u}{\ln\!\bigl(r/r_w\bigr)\, r_w^{2} K_w}
           + \dfrac{T_u}{T_l}
       \right)} 
\end{equation*}

Given inputs $\mathbf{x}_i$, the response is modeled as
\[
    y_{ij} \;=\; f(\mathbf{x}_i) \;+\; \varepsilon_{ij}, 
    \qquad \varepsilon_{ij} \stackrel{\text{iid}}{\sim} \mathcal{N}(0,1)
\]

Here $f(\mathbf{x}_i)$ is the deterministic borehole function that maps physical inputs to the idealized system response. In practice, repeated evaluations at the same inputs may differ due to measurement error. To account for this, we introduce the error term $\varepsilon_{ij}\sim\mathcal{N}(0,1)$. The zero mean ensures that noise does not bias the response away from $f(\mathbf{x}_i)$, while the unit variance ensures that the primary variability is explained by $f(\mathbf{x}_i)$ rather than by measurement noise. Thus, $\varepsilon_{ij}$ contributes only normalized random fluctuations around the underlying function.

For the borehole function with $p=8$ input variables, $n$ unique input locations were generated using Latin Hypercube (LH) sampling implemented in the \texttt{lhs} package \citep{Carnell2019lhs}, which provides a stratified design that distributes points evenly across the input space. Each variable was mapped from $[0,1]$ to its physical limits as summarized in Table~\ref{tab:borehole-inputs}. At each unique location, the deterministic borehole function $f(\mathbf{x}_i)$ was evaluated multiple times with added noise, resulting in a total of $N$ training observations. Test inputs were sampled uniformly at random and evaluated repeatedly with added noise.

\textbf{Simulation Setting:} We generate $n = 2000$ unique input locations, where each location is randomly replicated between one and three times, resulting in a total of $N = 3993$ noisy observations. Right-censoring is applied at three different thresholds: $C_1 = C_{0.90}$, $C_2 = C_{0.80}$, and $C_3 = C_{0.70}$, corresponding to the 90th, 80th, and 70th percentiles of the response distribution (approximately 143, 115, and 99, respectively).


{\small
\paragraph{Model Comparison and Censoring Scenarios}
To evaluate predictive performance under different censoring strategies, we consider four scenarios for comparison. In the \textit{Gold Standard} setting, no censored responses exist in the dataset. In contrast, under the \textit{Proposed Censored Likelihood Methods} setting, censored responses are present and are set according to a specified censoring threshold $C$.
In the \textit{Naive Approach 1} setting, censored responses are present but are fixed at the censoring threshold $C$ before model fitting. In the \textit{Naive Approach 2} setting, censored responses are removed entirely from the dataset prior to model fitting.

Based on these four scenarios, we conduct simulation studies to compare proposed methods with several baseline regression models. The proposed methods include \textit{SOV-Vecchia}, \textit{MC-SOV-Vecchia}, and \textit{MET-Vecchia}, which extend the Vecchia Gaussian process framework to handle censored data using different strategies for evaluating the censored likelihood. Specifically, \textit{SOV-Vecchia} employs deterministic integration through the separation of variables method, \textit{MC-SOV-Vecchia} applies Monte Carlo sampling, and \textit{MET-Vecchia} employs the minimax exponential tilting algorithm.  

For comparison, we also apply the Vecchia GP to the fully observed data and to the two naive censoring strategies. Additional benchmark models include \textit{laGP}, a local approximate Gaussian process \citep{gramacy2015local}; \textit{XGBoost}, the gradient boosting framework \citep{chen2016xgboost}; and conventional \textit{Linear} and \textit{Polynomial Regression} models.
}

{\small
\paragraph{Performance Metrics}
We evaluate model performance using multiple criteria. 
For predictive accuracy, we compute the mean squared prediction error (MSPE), mean absolute error (MAE), and mean absolute percentage error (MAPE). 
To assess predictive uncertainty, we use the average width of the 95\% predictive interval (Int), the empirical coverage rate of the 95\% interval (Cov), and the continuous ranked probability score (CRPS). 
We also compute tail prediction performance by reporting the percentage of test points where the predicted response correctly exceeds the censoring threshold among those with $y_i > C$ (\% Exceed $C$).
}

\subsection{Simulation Results}

We now present detailed results for the simulation setting to evaluate how the proposed censored-likelihood methods compare to baseline approaches across different censoring strategies.

Table~\ref{tab:20censored-pred} and Table~\ref{tab:20censored-uq} summarize the predictive accuracy and uncertainty quantification results when censoring occurs at the 80th percentile and the neighborhood size is set to $k = 20$. 
Results are reported separately for cases where the simulated noisy response satisfies $y < C$ and $y \ge C$, allowing evaluation of both the uncensored and censored regions.  

Under the \textit{Gold Standard}, no censoring scenario, where all responses are fully observed, the vGP model achieves the lowest prediction errors, smallest interval width (Int) and CRPS values across both regions.  

In contrast, under \textit{Naive Approach 1}, where censored values are fixed at the threshold $C$, the prediction accuracy decreases sharply when $y \ge C$. 
This happens because replacing censored data with $C$ introduces bias and prevents the model from learning the true pattern of the data. As a result, the model fails to capture the real behavior beyond the censoring limit. 
Similarly, under \textit{Naive Approach 2}, where censored values are removed from the dataset, the performance shows limited improvement. Removing censored observations reduces the amount of useful information available, which makes the predictions less stable. Overall, both naive methods produce wider intervals, lower coverage, and higher CRPS values when $y \ge C$.

For the \textit{Proposed Censored Likelihood Methods}, which explicitly account for censoring in the model, all three methods (SOV-Vecchia, MC-SOV-Vecchia, and MET-Vecchia) improve on the naive approaches in the censored region. 
SOV-Vecchia and MC-SOV-Vecchia substantially reduce prediction error in the censored region relative to the naive approaches. However, interval coverage above the censoring threshold remains
challenging, reflecting the difficulty of uncertainty quantification when the largest responses are only partially observed.
Overall, these results suggest that modeling censoring directly within the likelihood substantially improves point prediction and upper-tail classification relative to the naive approaches.

\begin{table}[t]
\centering
\footnotesize
\begin{tabular}{@{}lccc@{\hspace{7pt}\vrule\hspace{7pt}}ccc c@{}}
\toprule
\textbf{Model} &
\multicolumn{3}{c}{\textbf{ } $y<C$} &
\multicolumn{4}{c}{\textbf{ } $y\ge C$} \\ 
\cmidrule(lr){2-4}\cmidrule(lr){5-8}
 & MSPE & MAE & MAPE & MSPE & MAE & MAPE & \%Exceed$C$ \\
\midrule
\multicolumn{8}{c}{\textbf{Gold Standard: No Censoring}} \\
\midrule
vGP-Oracle & 1.53 & 0.99 & 2.33 & 1.50 & 0.97 & 0.65 & 100.0 \\
laGP       & 3.68 & 1.45 & 3.59 & 10.59 & 2.46 & 1.65 & 97.8 \\
XGBoost    & 11.47 & 2.63 & 5.53 & 29.12 & 4.38 & 2.99 & 95.7 \\
Linear     & 100.17 & 7.70 & 23.74 & 197.64 & 10.63 & 6.47 & 100.0 \\
Poly-2     & 69.75 & 5.83 & 18.14 & 123.63 & 8.60 & 5.51 & 100.0 \\
Poly-3     & 68.51 & 5.83 & 18.08 & 121.65 & 8.61 & 5.53 & 100.0 \\
\midrule
\multicolumn{8}{c}{\textbf{Proposed: Censored Likelihood Methods}} \\
\midrule
SOV-Vecchia     & 1.91 & 1.11 & 2.54 & 206.32 & 10.63 & 6.45 & 96.8 \\
MC-SOV-Vecchia  & 1.95 & 1.12 & 2.55 & 223.12 & 11.11 & 6.74 & 96.8 \\
MET-Vecchia     & 2.04 & 1.15 & 2.62 & 340.44 & 14.22 & 8.66 & 96.8 \\
\midrule
\multicolumn{8}{c}{\textbf{Naive Approach 1: Fixed at $C$}} \\
\midrule
vGP-Fixed  & 4.37 & 1.62 & 3.40 & 1425.00 & 31.20 & 19.28 & 44.3 \\
laGP       & 12.04 & 2.42 & 4.64 & 1412.12 & 31.35 & 19.51 & 46.6 \\
XGBoost    & 17.18 & 3.15 & 6.97 & 1564.17 & 34.14 & 21.49 & 36.6 \\
Linear     & 34.81 & 4.65 & 11.12 & 1156.07 & 30.83 & 19.77 & 42.7 \\
Poly-2     & 37.50 & 4.96 & 12.36 & 1258.59 & 32.20 & 20.63 & 40.8 \\
Poly-3     & 29.26 & 4.11 & 10.20 & 1385.48 & 33.42 & 21.32 & 37.3 \\
\midrule
\multicolumn{8}{c}{\textbf{Naive Approach 2: Censored Removed}} \\
\midrule
vGP-Removed & 3.63 & 1.51 & 3.68 & 475.54 & 17.72 & 11.35 & 58.2 \\
laGP        & 8.39 & 2.30 & 4.73 & 466.88 & 16.10 & 10.54 & 60.6 \\
XGBoost     & 12.52 & 2.48 & 4.53 & 1256.44 & 28.31 & 17.44 & 42.4 \\
Linear      & 34.09 & 4.56 & 12.58 & 896.51 & 26.70 & 17.02 & 48.5 \\
Poly-2      & 31.12 & 4.21 & 11.61 & 746.93 & 23.90 & 15.16 & 53.7 \\
Poly-3      & 27.67 & 3.93 & 10.79 & 977.66 & 27.81 & 17.68 & 44.1 \\
\bottomrule
\end{tabular}
\captionsetup{font=footnotesize}
\caption{Predictive performance under different censoring strategies.
The training set consists of $n = 2000$ unique input locations with a total of $N = 3993$ observations, 
and the test set includes $n = 200$ unique input locations with $N = 390$ observations. 
The number of nearest neighbors is set to $k = 20$. 
Censoring is applied at the 80th percentile, corresponding to a threshold of approximately $C \approx 115$.}
\label{tab:20censored-pred}
\end{table}

\begin{table}[t]
\centering
\footnotesize
\begin{tabular}{@{}lccc@{\hspace{7pt}\vrule\hspace{7pt}}ccc@{}}
\toprule
\textbf{Model} &
\multicolumn{3}{c}{\textbf{ } $y<C$} &
\multicolumn{3}{c}{\textbf{ } $y\ge C$} \\ 
\cmidrule(lr){2-4}\cmidrule(lr){5-7}
 & Int & Cov & CRPS & Int & Cov & CRPS \\
\midrule
\multicolumn{7}{c}{\textbf{Gold Standard: No Censoring}} \\
\midrule
vGP-Oracle & 5.0 & 96 & 0.70 & 5.1 & 97 & 0.69 \\
laGP       & 9.0 & 96 & 1.06 & 12.8 & 98 & 1.72 \\
XGBoost    & 8.2 & 79 & -- & 8.2 & 51 & -- \\
Linear     & 42.2 & 95 & 5.43 & 42.2 & 88 & 7.80 \\
Poly-2     & 36.0 & 95 & 4.53 & 36.0 & 88 & 6.19 \\
Poly-3     & 35.9 & 95 & 4.49 & 35.9 & 88 & 6.15 \\
\midrule
\multicolumn{7}{c}{\textbf{Proposed: Censored Likelihood Methods}} \\
\midrule
SOV-Vecchia     & 4.9 & 91 & 0.78 & 12.4 & 32 & 9.01 \\
MC-SOV-Vecchia  & 4.9 & 91 & 0.79 & 12.3 & 27 & 9.48 \\
MET-Vecchia     & 5.2 & 91 & 0.81 & 9.3 & 14 & 12.91 \\
\midrule
\multicolumn{7}{c}{\textbf{Naive Approach 1: Fixed at $C$}} \\
\midrule
vGP-Fixed  & 6.8 & 89 & 1.18 & 7.0 & 3 & 30.22 \\
laGP       & 7.9 & 82 & 1.81 & 11.8 & 3 & 29.66 \\
XGBoost    & 7.1 & 66 & -- & 7.1 & 1 & -- \\
Linear     & 26.9 & 97 & 3.33 & 26.9 & 10 & 27.01 \\
Poly-2     & 26.1 & 96 & 3.47 & 26.1 & 4 & 28.47 \\
Poly-3     & 22.6 & 95 & 3.02 & 22.6 & 4 & 30.18 \\
\midrule
\multicolumn{7}{c}{\textbf{Naive Approach 2: Censored Removed}} \\
\midrule
vGP-Removed & 6.3 & 89 & 1.09 & 12.1 & 13 & 11.22 \\
laGP        & 10.6 & 96 & 1.42 & 19.5 & 22 & 13.56 \\
XGBoost     & 3.8 & 54 & -- & 3.8 & 0 & -- \\
Linear      & 21.0 & 92 & 3.22 & 21.0 & 12 & 23.74 \\
Poly-2      & 20.4 & 91 & 3.05 & 20.4 & 15 & 21.08 \\
Poly-3      & 19.2 & 91 & 2.89 & 19.2 & 6 & 25.07 \\
\bottomrule
\end{tabular}
\captionsetup{font=footnotesize}
\caption{Uncertainty quantification performance under different censoring strategies. 
The training set consists of $n = 2000$ unique input locations with a total of $N = 3993$ observations, 
and the test set includes $n = 200$ unique input locations with $N = 390$ observations. 
The number of nearest neighbors is set to $k = 20$. 
Censoring is applied at the 80th percentile, corresponding to a threshold of approximately $C \approx 115$.}
\label{tab:20censored-uq}
\end{table}

We also evaluate the performance of the proposed censored likelihood GP model (SOV-Vecchia), which incorporates the censored likelihood through numerical integration using the SOV method. We compare its MSPE with that of the vecchia GP (vGP) fitted under the other three scenarios, Gold Standard, Naive Approach 1, and Naive Approach 2, across three censoring thresholds and two neighborhood sizes, $k = 10$ and $k = 20$.

\begin{figure}[H]
\centering
\includegraphics[width=1.0\textwidth, keepaspectratio]{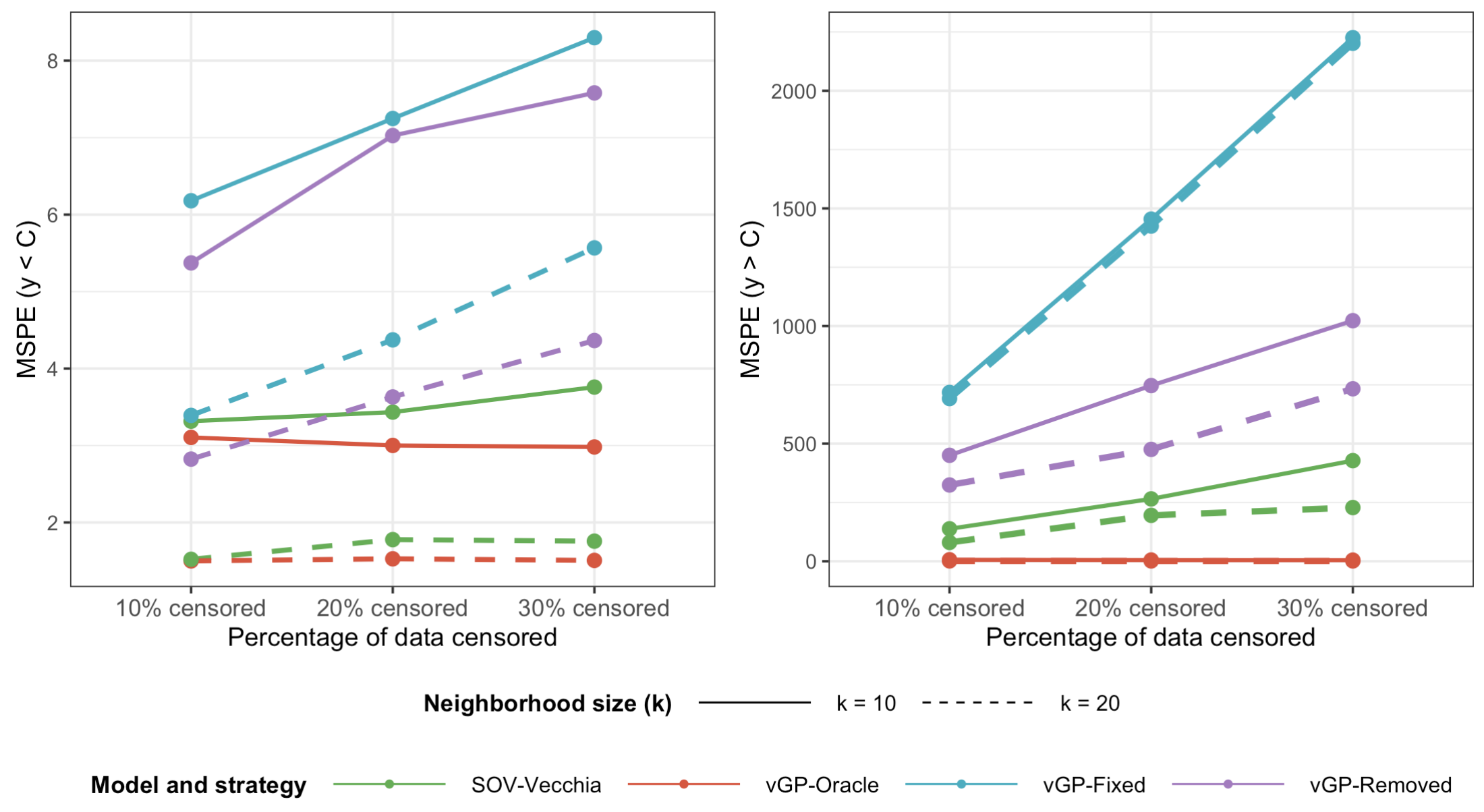}
\captionsetup{font=footnotesize}
\caption{
Comparison of prediction accuracy between the SOV-Vecchia and vGP under varying censoring levels and neighborhood sizes. 
Each panel shows the mean squared prediction error (MSPE) for 
(\textbf{left}) test cases with noisy responses below the censoring threshold ($y < C$) 
and (\textbf{right}) test cases with noisy responses above the threshold ($y \ge C$). 
Lines correspond to different strategies for handling censored data, 
and solid versus dashed lines indicate neighborhood sizes $k = 10$ and $k = 20$, respectively.
}
\label{fig:gp_compare}
\end{figure}

The left panel of Figure~\ref{fig:gp_compare} shows results for the case where noisy responses fall below the censoring threshold ($y < C$). The red line represents the vGP-Oracle model fitted to data without any censoring, which achieves the smallest mean squared prediction error (MSPE) and serves as the performance benchmark. The blue line corresponds to the vGP-Fixed model that treats censored observations as fixed at the detection threshold before fitting, and the purple line represents the vGP-Removed model fitted after removing censored observations. The green line shows the SOV-Vecchia model, which explicitly accounts for the censored likelihood. When the proportion of censored data increases, the MSPE rises sharply for the fixed-at-$C$ (blue) and removed-censored (purple) strategies. In contrast, SOV-Vecchia (green) remains consistently stable and improves as the neighborhood size increases from $k = 10$ to $k = 20$, achieving MSPE values comparable to the uncensored vGP-Oracle (red). 
This indicates that larger neighborhoods enable SOV-Vecchia to better capture local dependence structure, leading to improved predictive accuracy.

The right panel of Figure~\ref{fig:gp_compare} illustrates results for the case with noisy responses above the censoring threshold ($y \ge C$).
In this region, the vGP-Oracle model with no censoring (red line) again provides the lowest MSPE, while SOV-Vecchia (green line) maintains comparable performance, particularly as the neighborhood size increases to $k = 20$.
In contrast, the fixed-at-$C$ (blue line) and removed-censored (purple line) strategies show higher MSPE, as they cannot accurately model the upper tail of the response.
As more data are censored, the differences in performance between strategies become larger, highlighting the importance of properly handling censored observations. Overall, SOV-Vecchia remains relatively stable as the censoring proportion increases and improves as the neighborhood size increases from \(k=10\) to \(k=20\), remaining much closer to the oracle than the fixed-at-\(C\) and removed-censored strategies.

These results demonstrate that explicitly modeling censored likelihood through the SOV-Vecchia framework provides robust performance across varying censoring levels and remains substantially closer to the oracle model than the naive fixed-at-\(C\) and removed-censored strategies.

\section{Real Data Application}\label{sec: real}
\subsection{Data description}

We illustrate the performance of the proposed models using a real experimental dataset with a right-censored response.
This real-world dataset presents the challenges commonly encountered in industrial experiments with measurement replication to account for process variability, and censoring due to detection limits.
The response is subject to an upper detection limit such that any value exceeding a known threshold $C$ is recorded as censored. 
Approximately $46\%$ of the observations are censored, indicating that nearly half of the responses lie above the detection limit.
The experimental design involves 25 input variables ($\mathbf{x}_i \in \mathbb{R}^{25}$) representing process settings and environmental conditions. Replicated experiments are conducted at several distinct input locations to account for measurement variability. In total, the dataset contains $n = 1303$ unique input locations and $N = 6108$ observations, including both censored and uncensored responses. 
{\color{blue}
To assess the robustness of the real-data results, we repeated the random training and testing split 15 times. For each split, 90\% of the unique input locations were used for training and the remaining 10\% were used for testing.
Within each split, the 25 input variables were standardized using the training-set means and standard deviations, and the same transformations were applied to the testing inputs.}

\subsection{Results}

We evaluate the proposed censored-likelihood methods (MC-SOV-Vecchia and MET-Vecchia) against baseline approaches on this real dataset to assess their practical performance. Given that nearly half of the responses are censored, we focus on comparing methods' ability to both predict accurately in the uncensored region and extrapolate appropriately into the censored region.

For baseline methods that cannot handle censoring explicitly (laGP, XGBoost, linear regression, polynomial regression), we treat censored responses as fixed at the censoring threshold $C$ rather than removing them to avoid information loss. For the polynomial regression model, we restrict the functional form to second order given the 25-dimensional input space, as higher-order terms would substantially increase model complexity and risk overfitting.
{\color{blue}
To assess robustness, we repeated the real-data evaluation over multiple random splits, using 90\% of the data for training and 10\% for testing in each split. Table~\ref{tab:real_data} summarizes the performance across all methods, reporting the mean and standard deviation across repeated splits.
}
The evaluation metrics include prediction accuracy (MSPE, MAE) and uncertainty quantification (interval width, coverage, CRPS) computed on uncensored test points ($y < C$). Additionally, we report the percentage of predicted values exceeding the threshold (\% Exceed $C$) for test points with censored responses ($y \ge C$), which reflects each model's ability to extrapolate into the censored region.

\begin{table}[t]
\centering
\captionsetup{font=footnotesize}
{\footnotesize
\setlength{\tabcolsep}{6pt}
\renewcommand{\arraystretch}{1.05}
\begin{tabular}{l|ccccc|c}
\toprule
\textbf{Model} &
\multicolumn{5}{c|}{\textbf{$y < C$}} &
\multicolumn{1}{c}{\textbf{$y \ge C$}} \\
\cmidrule(lr){2-6} \cmidrule(lr){7-7}
 & \textbf{MSPE} & \textbf{MAE} & \textbf{Int} & \textbf{Cov} & \textbf{CRPS} &  
\textbf{\% Exceed $C$}\\ 
\midrule
MC-SOV-Vecchia & 2.22 (0.17) & 1.07 (0.04) & 6.89 (0.84) & 96.28 (1.53) & 0.81 (0.03) & 64.72 (2.89) \\ 
MET-Vecchia    & 2.34 (0.18) & 1.15 (0.09) & 7.37 (1.14) & 97.08 (1.52) & 0.87 (0.06) & 68.64 (4.98) \\ 
\midrule
laGP           & 3.99 (1.07) & 1.29 (0.13) & 11.28 (2.37) & 79.21 (5.36) & 1.21 (0.13) & 30.81 (6.33) \\ 
XGBoost        & 1.86 (0.24) & 1.02 (0.08) &  2.94 (0.03) & 76.36 (3.82) & --   & 26.69 (5.58) \\ 
Linear         & 2.18 (0.22) & 1.22 (0.07) & 5.34 (0.02) & 78.18 (2.27) & 0.85 (0.05) & 19.14 (2.19) \\ 
Poly-2         & 1.88 (0.21) & 1.13 (0.07) & 4.98 (0.02) & 70.95 (2.50) & 0.78 (0.05) & 22.92 (3.69) \\ 
\bottomrule
\end{tabular}
}
\caption{Summary of model performance on the real dataset. 
The dataset was split into 90\% for training and 10\% for testing. Values in parentheses are standard deviations across repeated splits.}
\label{tab:real_data}
\end{table}

{\color{blue}
The proposed censored-likelihood methods demonstrate strong overall performance. MC-SOV-Vecchia and MET-Vecchia achieve MSPE values of 2.22 and 2.34, respectively, with MAE values of 1.07 and 1.15. These results show competitive prediction accuracy in the uncensored region. More importantly, both methods provide substantially better uncertainty quantification than the baseline methods, with empirical coverage above 96\% and moderate interval widths. Their CRPS values are also low and stable across repeated splits, indicating reliable predictive distributions.

A key advantage of the proposed methods is their ability to extrapolate into the censored region. MC-SOV-Vecchia and MET-Vecchia achieve the highest percentages of predicted values above the censoring threshold, 64.72\% and 68.64\%, respectively. This indicates that explicitly modeling the censored likelihood makes the proposed methods more likely to predict responses above the censoring threshold for test cases known to be censored. Although the exact latent response values above \(C\) are unavailable, this metric suggests that the proposed methods preserve more upper-tail information than methods that fix censored observations at \(C\).

Among the baseline methods, XGBoost and the second-order polynomial model achieve low MSPE values, but both show poor uncertainty calibration and limited extrapolation into the censored region. XGBoost has narrow intervals and only 76.36\% coverage, while the polynomial model has 70.95\% coverage. Their \% Exceed $C$ values are also much lower than those of the proposed methods, suggesting that they tend to underpredict in the censored region.
The laGP method has larger prediction errors, with MSPE = 3.99, and substantially wider intervals, with average interval length 11.28. Although its intervals are wider, its empirical coverage remains only 79.21\%, and its \% Exceed $C$ is 30.81\%. This indicates that laGP does not adequately capture the censored upper tail in this application.
The linear and second-order polynomial regression models show moderate point-prediction accuracy but weak extrapolation ability, with \% Exceed $C$ values of only 19.14\% and 22.92\%, respectively. This suggests that simple parametric forms cannot adequately capture the nonlinear relationship between the 25-dimensional inputs and the response, particularly in the upper tail where censoring occurs.

Overall, these real-data results corroborate the simulation findings in Section~\ref{sec: simu}. Explicitly accounting for censoring through likelihood-based methods improves uncertainty quantification and substantially improves extrapolation to censored regions. The proposed MC-SOV-Vecchia and MET-Vecchia methods are able to use information from censored observations while maintaining computational feasibility for this moderately large dataset ($n = 1303$ unique input locations, $N = 6108$ total observations).
}

{\color{blue}
We also report the actual computation times of the three censored-probability evaluation methods, which differ in how they evaluate the censored Gaussian tail probability. SOV uses direct numerical integration and is efficient only for very low-dimensional censored blocks (block dimension between one and three in our implementation); MC--SOV uses a Monte Carlo approximation based on the separation-of-variables transformation and handles larger blocks; and MET additionally solves a local minimax tilting optimization, making it the most expensive. For the real dataset, some censored blocks have dimension larger than three, so direct SOV was not used; with Vecchia neighborhood size $k=20$ and Monte Carlo sample size $M=50{,}000$, the average computation times were $6678$ seconds for MC--SOV and $8109$ seconds for MET. For the eight-dimensional Borehole simulation ($n=2000$ unique input locations, $N=3993$ observations, replicate sizes between one and three, and about $20\%$ censored responses), all three methods were applicable, with computation times of $1312$, $4837$, and $6601$ seconds for SOV, MC--SOV, and MET, respectively (Table~\ref{tab:runtime_methods}).

\begin{table}[t]
\centering
\caption{Computation times for the censored-probability evaluation methods, under Vecchia neighborhood size $k=20$.}
\label{tab:runtime_methods}
\begin{tabular}{lccc}
\toprule
Setting & SOV & MC--SOV & MET \\
\midrule
Real data & Not used & 6678 s & 8109 s \\
Borehole simulation & 1312 s & 4837 s & 6601 s \\
\bottomrule
\end{tabular}
\end{table}
}

\section{Discussion}\label{sec: disc}
This paper introduces a scalable Gaussian process framework for modeling datasets with both replicated and censored observations.
By integrating Vecchia-based approximations with likelihood-based methods for censored data, the proposed approach improves computational feasibility while preserving information from
censored observations. The numerical studies show that explicitly accounting for censoring improves upper-tail prediction and uncertainty calibration relative to common non-censoring-aware strategies, although uncertainty quantification above the censoring threshold remains challenging when censoring is severe.
{\color{blue}
The additional sensitivity analysis in Appendix~\ref{app:unbalanced_noise} further examines the robustness of the proposed method under highly unbalanced
replication and larger noise variance. The results show that the method remains stable when some input locations have many replicates while others have only one replicate. Compared with the balanced sparse setting, the prediction performance in the uncensored region remains similar, while the performance in the censored region improves because the additional replicated observations provide useful information. When the noise variance is increased, prediction becomes more difficult, as expected, leading to larger point prediction errors and wider predictive intervals. However, the coverage remains reasonable, especially in
the censored region. These findings suggest that the proposed framework is applicable beyond the basic Borehole setting and can handle more realistic industrial data structures with uneven replication and increased measurement noise.
}

The current framework handles right-censored observations and can be readily extended to interval censoring, where responses are known only to lie within an interval $[C_L, C_U]$. This extension would enable the model to handle more complex measurement limitations, such as data falling below detection limits or between multiple thresholds. For datasets with a large proportion of censored observations, the censored probability evaluation can become computationally intensive. In such cases, the framework can leverage the efficient algorithms developed by \citet{cao_katzfuss_2025_vecchia} for evaluating high-dimensional truncated multivariate normal densities, further improving scalability.

\textcolor{blue}{Motivated by the multi-output censored Gaussian process of \citet{gammelli2022generalized},} the framework can be combined with models for multivariate outputs, where multiple correlated responses are observed at each input location, enabling joint modeling of complex systems with multiple quality characteristics or performance metrics. For problems with high-dimensional inputs or outputs, the framework can be integrated with dimension reduction techniques to identify the most important features and improve computational efficiency. Such extensions would broaden the applicability of the method to modern high-dimensional experimental settings common in manufacturing, materials science, and other engineering applications. Finally, while the present work focuses on inference and prediction under replication and censoring, future research may explore the use of the proposed censoring-aware Gaussian process framework within sequential experimental design, where accurate uncertainty quantification under incomplete observations is critical.

\section*{Data Availability Statement}

The synthetic data and code supporting the findings presented in Section~\ref{sec: simu} are publicly available in the GitHub repository. The real data analyzed in Section~\ref{sec: real} are not publicly available due to the nature of the research.

\bibliographystyle{plainnat}   
\bibliography{references}      

\appendix

\section{Derivation of Proposition~\ref{prop:ynew_cens}}
\label{app:prediction_derivation}
\paragraph{Step A.1: Joint Gaussian structure}
Under the Gaussian process model, the joint vector
$
\bigl(y_{\mathrm{new}}, \mathbf y_c, \mathbf y_o\bigr)
$ is multivariate normal. We begin by conditioning on the uncensored neighbors $\mathbf y_o$ using standard Gaussian conditioning formulas. The joint conditional distribution of $y_{\mathrm{new}}$ and latent variable $\mathbf y_c$ given $\mathbf y_o$ is multivariate normal:
\begin{equation*}
\begin{pmatrix}
y_{\mathrm{new}} \\
\mathbf y_c
\end{pmatrix}
\Big| \mathbf y_o
\sim
\mathcal N\!\left(
\begin{pmatrix}
\mu_{\mathrm{new}\mid o} \\
\boldsymbol\mu_{c\mid o}
\end{pmatrix},
\begin{pmatrix}
\boldsymbol\Sigma_{\mathrm{new}\mid o} &
\boldsymbol\Sigma_{\mathrm{new},c\mid o} \\
\boldsymbol\Sigma_{c,\mathrm{new}\mid o} &
\boldsymbol\Sigma_{cc\mid o}
\end{pmatrix}
\right)
\label{eq:joint_conditional}
\end{equation*}

\paragraph{Step A.2: Conditional distribution given $(\mathbf y_o, \mathbf y_c)$ }
From the joint Gaussian structure in Step A.1, if $\mathbf y_c$ were fully observed, the conditional distribution would be:
\begin{equation}
y_{\mathrm{new}} \mid \mathbf y_o, \mathbf y_c
\sim
\mathcal N\!\left(
\boldsymbol\mu_{\mathrm{new}\mid o}
+
\boldsymbol\Sigma_{\mathrm{new},c\mid o}
\boldsymbol\Sigma_{cc\mid o}^{-1}
(\mathbf y_c-\boldsymbol\mu_{c\mid o}),
\;
\boldsymbol\Sigma_{\mathrm{new}\mid o}
-
\boldsymbol\Sigma_{\mathrm{new},c\mid o}
\boldsymbol\Sigma_{cc\mid o}^{-1}
\boldsymbol\Sigma_{c,\mathrm{new}\mid o}
\right)
\label{eq:conditional_given_yc}
\end{equation}

\paragraph{Step A.3: Conditioning on the censoring}
\textcolor{black}{The latent vector \(\mathbf y_c\) is not observed directly; only the orthant event \(\{\mathbf y_c\ge\mathbf C\}\) is observed.} Therefore, the conditional distribution under censoring is obtained by integrating over the latent $\mathbf{y}_c$ in the region $\{\mathbf{y}_c \ge \mathbf{C}\}$:
\begin{equation}
p(y_{\mathrm{new}} \mid \mathbf y_o, \mathbf y_c \ge \mathbf C)
=
\int
p(y_{\mathrm{new}} \mid \mathbf y_o, \mathbf y_c)\,
p(\mathbf y_c \mid \mathbf y_o, \mathbf y_c \ge \mathbf C)
\, d\mathbf y_c
\label{eq:integration}
\end{equation}

\textcolor{black}{This integral defines a closed skew-normal distribution in general, rather than a Gaussian distribution. The predictive mean and variance are derived in Steps A.4 and A.5, and an exact composition sampler is provided in Appendix~\ref{app:prediction_sampling}.}

\paragraph{Step A.4: Predictive mean under censoring}
\textcolor{black}{Starting from Step A.3 and applying the law of total expectation,
the predictive mean under censoring is}
\begin{equation}
\mathbb{E}\!\left( y_{\mathrm{new}} \mid \mathbf y_o, \mathbf y_c \ge \mathbf C \right)
=
\int
\mathbb{E}\!\left( y_{\mathrm{new}} \mid \mathbf y_o, \mathbf y_c \right)
\, p\!\left( \mathbf y_c \mid \mathbf y_o, \mathbf y_c \ge \mathbf C \right)
\, d\mathbf y_c 
\label{eq:integral}
\end{equation}

From Step A.2, if $\mathbf{y}_c$ were observed, the conditional expectation would be:
\begin{equation*}
\mathbb{E}\!\left( y_{\mathrm{new}} \mid \mathbf y_o, \mathbf y_c \right)
=
\boldsymbol\mu_{\mathrm{new}\mid o}
+
\boldsymbol{\Sigma}_{\mathrm{new},c\mid o}
\boldsymbol{\Sigma}^{-1}_{cc\mid o}
\left( \mathbf y_c - \boldsymbol{\mu}_{c\mid o} \right)
\end{equation*}

Substituting this expression into the integral \eqref{eq:integral}  
\begin{align}
\mathbb{E}(y_{\text{new}} \mid \mathbf{y}_o, \mathbf{y}_c \ge \mathbf{C}) 
&= \int \left[\boldsymbol{\mu}_{\text{new}|o} + \boldsymbol\Sigma_{\text{new},c|o}\boldsymbol\Sigma_{cc|o}^{-1} (\mathbf{y}_c - \boldsymbol{\mu}_{c|o})\right] p(\mathbf{y}_c \mid \mathbf{y}_o, \mathbf{y}_c \ge \mathbf{C}) \, d\mathbf{y}_c \nonumber \\
&= \boldsymbol{\mu}_{\text{new}|o} \int p(\mathbf{y}_c \mid \mathbf{y}_o, \mathbf{y}_c \ge \mathbf{C}) \, d\mathbf{y}_c \nonumber \\
&\quad + \boldsymbol\Sigma_{\text{new},c|o}\boldsymbol\Sigma_{cc|o}^{-1} \int (\mathbf{y}_c - \boldsymbol{\mu}_{c|o}) \, p(\mathbf{y}_c \mid \mathbf{y}_o, \mathbf{y}_c \ge \mathbf{C}) \, d\mathbf{y}_c \nonumber \\
&= \boldsymbol{\mu}_{\text{new}|o} + \boldsymbol\Sigma_{\text{new},c|o}\boldsymbol\Sigma_{cc|o}^{-1} \left[\int \mathbf{y}_c \, p(\mathbf{y}_c \mid \mathbf{y}_o, \mathbf{y}_c \ge \mathbf{C}) \, d\mathbf{y}_c - \boldsymbol{\mu}_{c|o}\right] \nonumber \\
&= \boldsymbol{\mu}_{\text{new}|o} + \boldsymbol\Sigma_{\text{new},c|o}\boldsymbol\Sigma_{cc|o}^{-1} \left(\mathbb{E}(\mathbf{y}_c \mid \mathbf{y}_o, \mathbf{y}_c \ge \mathbf{C}) - \boldsymbol{\mu}_{c|o}\right) 
\label{eq:mu_star}
\end{align}

Define
\[
\mathbf m_c
=
\mathbb E(\mathbf y_c \mid \mathbf y_o, \mathbf y_c\ge\mathbf C)
\]
Substituting into \eqref{eq:mu_star} gives
\begin{equation*}
\boldsymbol\mu_{\mathrm{new}}^*
=
\boldsymbol\mu_{\mathrm{new}\mid o}
+
\boldsymbol\Sigma_{\mathrm{new},c\mid o}
\boldsymbol\Sigma_{cc\mid o}^{-1}
(\mathbf m_c-\boldsymbol\mu_{c\mid o})
\end{equation*}
which establishes \eqref{eq:mu_star_formal}.

\paragraph{Step A.5: Predictive variance under censoring}
The predictive variance under censoring is obtained using the law of total variance:
\begin{align}
\mathrm{Var}(y_{\mathrm{new}} \mid \mathbf y_o, \mathbf y_c\ge\mathbf C)
&=
\mathbb E\!\left[
\mathrm{Var}(y_{\mathrm{new}} \mid \mathbf y_o, \mathbf y_c)
\mid \mathbf y_o, \mathbf y_c\ge\mathbf C
\right] \notag \\
&\quad+
\mathrm{Var}\!\left[
\mathbb E(y_{\mathrm{new}} \mid \mathbf y_o, \mathbf y_c)
\mid \mathbf y_o, \mathbf y_c\ge\mathbf C
\right]
\label{eq:totvar}
\end{align}

For the first term, since the conditional variance from Step A.2 does not depend on $\mathbf{y}_c$, taking the expectation with respect to the truncated distribution gives the same expression:

\begin{equation*}
\mathbb{E}[\text{Var}(y_{\text{new}} \mid \mathbf{y}_o, \mathbf{y}_c) \mid \mathbf{y}_o, \mathbf{y}_c \ge \mathbf{C}]=
\boldsymbol\Sigma_{\mathrm{new}\mid o}
-
\boldsymbol\Sigma_{\mathrm{new},c\mid o}
\boldsymbol\Sigma_{cc\mid o}^{-1}
\boldsymbol\Sigma_{c,\mathrm{new}\mid o}
\end{equation*}

For the second term, we evaluate the variance of the conditional mean with respect to the truncated distribution of $\mathbf y_c$
\begin{align*}
&\text{Var}[\mathbb{E}(y_{\text{new}} \mid \mathbf{y}_o, \mathbf{y}_c) \mid \mathbf{y}_o, \mathbf{y}_c \ge \mathbf{C}] \nonumber \\
&= \text{Var}\left[\boldsymbol{\mu}_{\text{new}|o} + \boldsymbol\Sigma_{\text{new},c|o}\boldsymbol\Sigma_{cc|o}^{-1} (\mathbf{y}_c - \boldsymbol{\mu}_{c|o}) \mid \mathbf{y}_o, \mathbf{y}_c \ge \mathbf{C}\right] \nonumber \\
&= \text{Var}\left[\boldsymbol\Sigma_{\text{new},c|o}\boldsymbol\Sigma_{cc|o}^{-1} (\mathbf{y}_c - \boldsymbol{\mu}_{c|o}) \mid \mathbf{y}_o, \mathbf{y}_c \ge \mathbf{C}\right] \nonumber \\
&= \boldsymbol\Sigma_{\text{new},c|o}\boldsymbol\Sigma_{cc|o}^{-1} \, \text{Var}(\mathbf{y}_c - \boldsymbol{\mu}_{c|o} \mid \mathbf{y}_o, \mathbf{y}_c \ge \mathbf{C}) \, \boldsymbol\Sigma_{cc|o}^{-1}\boldsymbol\Sigma_{c,\text{new}|o} \nonumber \\
&= \boldsymbol\Sigma_{\mathrm{new},c\mid o}
\boldsymbol\Sigma_{cc\mid o}^{-1}
\,\mathrm{Var}(\mathbf y_c \mid \mathbf y_o, \mathbf y_c\ge\mathbf C)\,
\boldsymbol\Sigma_{cc\mid o}^{-1}
\boldsymbol\Sigma_{c,\mathrm{new}\mid o}
\end{align*}

Define
\[
\mathbf V_c
=
\mathrm{Var}(\mathbf y_c \mid \mathbf y_o, \mathbf y_c\ge\mathbf C)
\]
Combining two terms yields
\begin{equation*}
\boldsymbol\Sigma_{\mathrm{new}}^*
=
\boldsymbol\Sigma_{\mathrm{new}\mid o}
-
\boldsymbol\Sigma_{\mathrm{new},c\mid o}
\boldsymbol\Sigma_{cc\mid o}^{-1}
\boldsymbol\Sigma_{c,\mathrm{new}\mid o}
+
\boldsymbol\Sigma_{\mathrm{new},c\mid o}
\boldsymbol\Sigma_{cc\mid o}^{-1}
\mathbf V_c
\boldsymbol\Sigma_{cc\mid o}^{-1}
\boldsymbol\Sigma_{c,\mathrm{new}\mid o}
\end{equation*}
which establishes \eqref{eq:Sig_star_formal}.

\section{Computation of truncated moments}
\label{app:prediction_moment}
To evaluate the predictive mean and variance in
\eqref{eq:mu_star_formal} and \eqref{eq:Sig_star_formal}, we require the
truncated moments
\[
\mathbf m_c
=
\mathbb E(\mathbf y_c \mid \mathbf y_o, \mathbf y_c\ge\mathbf C),
\qquad
\mathbf V_c
=
\mathrm{Var}(\mathbf y_c \mid \mathbf y_o, \mathbf y_c\ge\mathbf C)
\]

From the joint Gaussian structure, the conditional distribution of the latent censored responses given the observed data is
\[
\mathbf y_c \mid \mathbf y_o
\sim
\mathcal N(\boldsymbol\mu_{c\mid o}, \boldsymbol\Sigma_{cc\mid o})
\]
Conditioning this distribution on the $\mathbf y_c\ge\mathbf C$ yields a truncated multivariate normal distribution supported on the region $\{\mathbf y_c\ge\mathbf C\}$. 
The moments $\mathbf m_c$ and $\mathbf V_c$ are defined with respect to this truncated distribution and are computed numerically. 
In this work, we employ the \texttt{tmvtnorm} package  \citep{WilhelmManjunath2015} in R, which evaluates these moments directly for general covariance structures and truncation regions using the algorithm of \citet{GenzBretz2009}.


\subsection*{B.1 Standardization via Cholesky decomposition}

Let $\boldsymbol\Sigma_{cc\mid o}=\mathbf L\mathbf L^\top$ be the Cholesky factorization.
Define the standardized variable
\[
\mathbf z = \mathbf L^{-1}(\mathbf y_c-\boldsymbol\mu_{c\mid o}),
\]
so that $\mathbf z\sim\mathcal N(\mathbf 0,\mathbf I)$.

Writing $\mathbf z_{1:(j-1)}=(z_1,\ldots,z_{j-1})^\top$, the sequential truncation bounds are
\[
\alpha_j(\mathbf z_{1:(j-1)})
=
\frac{C_j-\mu_{c\mid o,j}-\sum_{k<j}L_{jk}z_k}{L_{jj}},
\qquad
j=1,\ldots,d_c,
\]
where $d_c=\dim(\mathbf y_c)$, and the bounds for component $j$ depend on the values of previous components through the Cholesky factor $\mathbf{L}$.

Let
\[
\mathbf m_z=\E(\mathbf z\mid \mathbf z\ge\boldsymbol\alpha),
\qquad
\mathbf V_z=\Var(\mathbf z\mid \mathbf z\ge\boldsymbol\alpha)
\]
These standardized moments are computed using the \texttt{tmvtnorm} package in R, which implements efficient algorithms for truncated normal moments.

\subsection*{B.2 Transformation back to the original scale}

Given estimates $\mathbf m_z$ and $\mathbf V_z$,
the truncated moments of $\mathbf y_c$ are obtained via
\[
\E(\mathbf y_c\mid \mathbf y_o,\mathbf y_c\ge\mathbf C)
=
\boldsymbol\mu_{c\mid o}+\mathbf L\mathbf m_z,
\qquad
\Cov(\mathbf y_c\mid \mathbf y_o,\mathbf y_c\ge\mathbf C)
=
\mathbf L\mathbf V_z\mathbf L^\top
\]
These quantities are substituted into
\eqref{eq:mu_star_formal}--\eqref{eq:Sig_star_formal}
to obtain the predictive mean and variance of $y_{\new}$ under censoring.

\section{Sampling from the predictive distribution}
\label{app:prediction_sampling}
{\color{black}

The predictive distribution under censoring is generally non-Gaussian and belongs to the closed skew-normal family
\citep{DominguezMolinaEtAl2003,GonzalezFariasEtAl2004}. Although its predictive mean and variance are available analytically from
Proposition~\ref{prop:ynew_cens}, these moments alone are insufficient for constructing prediction intervals. The mixture representation in \eqref{eq:integration} leads directly to a two-stage composition sampling procedure: first sample the latent censored responses from their truncated conditional distribution, and then sample the new response conditional on the sampled censored responses.

For \(b=1,\ldots,B\), perform the following steps:

\paragraph{Step 1: Sample the latent censored responses}

Conditional on the observed uncensored responses,
\[
\mathbf y_c\mid\mathbf y_o
\sim
\mathcal N\!\left(
\boldsymbol\mu_{c\mid o},
\boldsymbol\Sigma_{cc\mid o}
\right)
\]
Because the latent censored responses are known to satisfy
\(\mathbf y_c\ge\mathbf C\), we generate
\[
\mathbf y_c^{(b)}
\sim
\mathcal N\!\left(
\boldsymbol\mu_{c\mid o},
\boldsymbol\Sigma_{cc\mid o}
\right)
\quad\text{truncated to }
\{\mathbf y_c \ge \mathbf C\}
\]

Each draw represents a possible realization of the latent censored responses that is consistent with the observed uncensored responses and the censoring constraints. In our implementation, the truncated multivariate normal draws are generated using the exact minimax-tilting method of \citet{botev2017normal}, as implemented in the \texttt{TruncatedNormal} package in R.

\paragraph{Step 2: Sample the new response}

For the \(b\)th Monte Carlo iteration, the draw \(\mathbf y_c^{(b)}\) provides one possible realization of the latent
censored responses. Conditional on this realization and the observed uncensored responses \(\mathbf y_o\), the new response follows the Gaussian conditional distribution in \eqref{eq:conditional_given_yc}. We therefore generate
\[
y_{\new}^{(b)}
\sim
\mathcal N\!\left(
\boldsymbol\mu_{\new\mid o}
+
\boldsymbol\Sigma_{\new,c\mid o}
\boldsymbol\Sigma_{cc\mid o}^{-1}
\left(
\mathbf y_c^{(b)}-\boldsymbol\mu_{c\mid o}
\right),
\;
\boldsymbol\Sigma_{\new\mid o}
-
\boldsymbol\Sigma_{\new,c\mid o}
\boldsymbol\Sigma_{cc\mid o}^{-1}
\boldsymbol\Sigma_{c,\new\mid o}
\right)
\]

The distribution of \(y_{\new}\mid\mathbf y_o,\mathbf y_c\ge\mathbf C\) is generally non-Gaussian, with predictive mean and variance given by \eqref{eq:mu_star_formal} and \eqref{eq:Sig_star_formal}.
A \(100(1-\alpha)\%\) prediction interval is constructed from the empirical \(\alpha/2\) and \(1-\alpha/2\) quantiles of
\[
y_{\new}^{(1)},\ldots,y_{\new}^{(B)}
\]
These intervals are generally asymmetric because of the skewness induced by censoring.

}

\section{Inverse-transform sampling in MC--SOV}
\label{app:sov_inverse}

In the MC--SOV algorithm, each component \(z_{i,j}\) is sampled sequentially
from a standard normal distribution truncated below at
\(\alpha_j(z_{i,1},\ldots,z_{i,j-1})\). Conditional on the previously sampled
components \(z_{i,1},\ldots,z_{i,j-1}\), its distribution function is
\[
F_{j}(t\mid z_{i,1:j-1}) \;=\;
\frac{\Phi(t)-\Phi\!\big(\alpha_{j}(z_{i,1:j-1})\big)}
     {1-\Phi\!\big(\alpha_{j}(z_{i,1:j-1})\big)},
\qquad 
t\in\big[\alpha_{j}(z_{i,1:j-1}),\,\infty\big]
\]
Applying the inverse transform sampling theorem to the conditional CDF, 
we take $U_{j}\sim\mathrm{Unif}(0,1)$ and set $U_{j} = F_{j}(z_{i,j}\mid z_{i,1:j-1})$. 
Solving for $z_{i,j}$ gives:
\begin{align*}
U_{j}[1-\Phi(\alpha_{j})] &= \Phi(z_{i,j})-\Phi(\alpha_{j}) \\[3pt]
\Phi(z_{i,j}) &= \Phi(\alpha_{j}) + U_{i,j}[1-\Phi(\alpha_{j})] \\[3pt]
z_{i,j} &= \Phi^{-1}\!\Big(\Phi(\alpha_{j}) + U_{j}[1-\Phi(\alpha_{j})]\Big)
\end{align*}

Thus, for $j = 1, \ldots, r_i-m_i$, the uniform random variable 
$U_{j} \sim U(0,1)$ is generated and transformed through the inverse of the standard normal CDF set as
\[
z_{i,j} = \Phi^{-1}\!\left[\Phi\!\big(\alpha_{j}(z_{i,1}, \ldots, z_{i,j-1})\big)
+ U_{j}\!\left\{1
- \Phi\!\big(\alpha_{j}(z_{i,1}, \ldots, z_{i,j-1})\big)\right\}\right],
\]
where $\Phi(\cdot)$ denotes the standard normal cumulative distribution function.
This is the sampling formula used in the MC--SOV algorithm.

\section{Minimax exponential tilting details}
\label{app:met_details}

The standard SOV estimator samples sequentially from truncated standard normal distributions. When the target probability is small, this estimator can be inefficient because the resulting
Monte Carlo weights may be unstable. MET improves the SOV proposal by introducing a tilting
parameter
\[
\boldsymbol{\lambda}_i
=
(\lambda_{i,1},\ldots,\lambda_{i,d_i})^\top 
\]
For each component \(j\), the tilted conditional proposal is
\[
g_j
\left(
z_{i,j}\mid z_{i,1},\ldots,z_{i,j-1};
\boldsymbol{\lambda}_i
\right)
=
\frac{
\phi(z_{i,j};\lambda_{i,j},1)
I
\left\{
z_{i,j}\ge
\alpha_j(z_{i,1},\ldots,z_{i,j-1})
\right\}
}{
1-\Phi\left[
\alpha_j(z_{i,1},\ldots,z_{i,j-1})-\lambda_{i,j}
\right]
}
\]
Thus, the joint tilted SOV proposal density is
\[
g_{\boldsymbol{\lambda}_i}(\mathbf{z}_i)
=
\prod_{j=1}^{d_i}
g_j
\left(
z_{i,j}\mid z_{i,1},\ldots,z_{i,j-1};
\boldsymbol{\lambda}_i
\right),
\qquad
\mathbf{z}_i\in\mathcal{R}_i 
\]

Under this proposal, the importance weight is the likelihood ratio between the standard normal integrand and the tilted SOV proposal
\[
w(\mathbf{z}_i;\boldsymbol{\lambda}_i)
=
\frac{
\phi(\mathbf{z}_i;\mathbf{0},\mathbf{I}_{d_i})
}{
g_{\boldsymbol{\lambda}_i}(\mathbf{z}_i)
}
=
\exp
\left\{
\psi_i(\mathbf{z}_i,\boldsymbol{\lambda}_i)
\right\},
\]
where
\[
\psi_i(\mathbf{z}_i,\boldsymbol{\lambda}_i)
=
-\mathbf{z}_i^\top\boldsymbol{\lambda}_i
+
\frac{1}{2}\|\boldsymbol{\lambda}_i\|^2
+
\sum_{j=1}^{d_i}
\log
\left[
1-\Phi
\left\{
\alpha_j(z_{i,1},\ldots,z_{i,j-1})-\lambda_{i,j}
\right\}
\right]
\]
Therefore,
\[
p_i
=
E_{g_{\boldsymbol{\lambda}_i}}
\left[
\exp
\left\{
\psi_i(\mathbf{Z}_i,\boldsymbol{\lambda}_i)
\right\}
\right]
\]

The choice of \(\boldsymbol{\lambda}_i\) determines the variance of the estimator. Following \citet{botev2017normal}, the tilting parameter is chosen by controlling the worst-case likelihood
ratio over the truncation region. Specifically,
\[
\boldsymbol{\lambda}_i^*
=
\arg\min_{\boldsymbol{\lambda}_i}
\sup_{\mathbf{z}_i\in\mathcal{R}_i}
\psi_i(\mathbf{z}_i,\boldsymbol{\lambda}_i)
\]
Equivalently, by the minimax saddle-point formulation of \citet{botev2017normal}, the optimal pair \((\mathbf{z}_i^*,\boldsymbol{\lambda}_i^*)\) can be characterized as
\[
(\mathbf{z}_i^*,\boldsymbol{\lambda}_i^*)
=
\arg\max_{\mathbf{z}_i,\boldsymbol{\lambda}_i}
\psi_i(\mathbf{z}_i,\boldsymbol{\lambda}_i),
\qquad
\text{subject to }
\frac{\partial \psi_i}{\partial \boldsymbol{\lambda}_i}
=
\mathbf{0},
\quad
\mathbf{z}_i\in\mathcal{R}_i 
\]

To write the stationarity equations explicitly, define
\[
q_j
=
1-\Phi
\left\{
\alpha_j(z_{i,1},\ldots,z_{i,j-1})-\lambda_{i,j}
\right\},
\]
and
\[
\rho_j
=
\frac{
\phi
\left\{
\alpha_j(z_{i,1},\ldots,z_{i,j-1})-\lambda_{i,j}
\right\}
}{
q_j
}
\]
Then
\[
\frac{\partial \psi_i}{\partial \lambda_{i,j}}
=
-z_{i,j}
+
\lambda_{i,j}
+
\rho_j,
\qquad j=1,\ldots,d_i,
\]
and
\[
\frac{\partial \psi_i}{\partial z_{i,j}}
=
-\lambda_{i,j}
+
\sum_{\ell=j+1}^{d_i}
\frac{L_{i,\ell j}}{L_{i,\ell\ell}}\rho_\ell,
\qquad j=1,\ldots,d_i
\]
The unconstrained stationary equations are
\[
-z_{i,j}
+
\lambda_{i,j}
+
\rho_j
=
0,
\qquad j=1,\ldots,d_i,
\]
and
\[
-\lambda_{i,j}
+
\sum_{\ell=j+1}^{d_i}
\frac{L_{i,\ell j}}{L_{i,\ell\ell}}\rho_\ell
=
0,
\qquad j=1,\ldots,d_i
\]
Equivalently,
\[
z_{i,j}
=
\lambda_{i,j}
+
\rho_j,
\qquad
\lambda_{i,j}
=
\sum_{\ell=j+1}^{d_i}
\frac{L_{i,\ell j}}{L_{i,\ell\ell}}\rho_\ell 
\]

Following the practical implementation of \citet{botev2017normal}, these equations are first solved as an unconstrained problem using Powell's trust-region dogleg method \citep{powell1970optimization}, giving a candidate solution
$
(\widetilde{\mathbf{z}}_i,\widetilde{\boldsymbol{\lambda}}_i).
$
If \(\widetilde{\mathbf{z}}_i\in\mathcal{R}_i\), we set \(\boldsymbol{\lambda}_i^* = \widetilde{\boldsymbol{\lambda}}_i\). If instead the candidate is infeasible or the trust-region method fails to converge, the constrained saddle-point problem is solved directly over \(\mathbf{z}_i\in\mathcal{R}_i\), using \((\widetilde{\mathbf{z}}_i,\widetilde{\boldsymbol{\lambda}}_i)\) as the initial value. Either branch gives the optimal tilting parameter \(\boldsymbol{\lambda}_i^*\). 
Given \(\boldsymbol{\lambda}_i^*\), samples are generated sequentially from the tilted truncated proposal \(g_{\boldsymbol{\lambda}_i^*}\). Specifically, for \(m=1,\ldots,M\) and \(j=1,\ldots,d_i\), draw \(U_{i,j}^{(m)}\sim \mathrm{Unif}(0,1)\) and set
\[
z_{i,j}^{(m)}
=
\lambda_{i,j}^*
+
\Phi^{-1}
\left[
\Phi
\left\{
\alpha_j(z_{i,1}^{(m)},\ldots,z_{i,j-1}^{(m)})
-
\lambda_{i,j}^*
\right\}
+
U_{i,j}^{(m)}
\left\{
1-
\Phi
\left[
\alpha_j(z_{i,1}^{(m)},\ldots,z_{i,j-1}^{(m)})
-
\lambda_{i,j}^*
\right]
\right\}
\right]
\]
This sequential inverse-transform step generates
\[
\mathbf{z}_i^{(m)}
\sim
g_{\boldsymbol{\lambda}_i^*},
\qquad
\mathbf{z}_i^{(m)}\in\mathcal{R}_i 
\]

For a generic draw \(\mathbf{z}_i\in\mathcal{R}_i\), the importance weight is
\[
w(\mathbf{z}_i)
=
\frac{
\phi(\mathbf{z}_i;\mathbf{0},\mathbf{I}_{d_i})
}{
g_{\boldsymbol{\lambda}_i^*}(\mathbf{z}_i)
}
\]
Using the product form of the tilted SOV proposal, this ratio can be written as
\[
w(\mathbf{z}_i)
=
\exp
\left\{
-\mathbf{z}_i^\top\boldsymbol{\lambda}_i^*
+
\frac{1}{2}\|\boldsymbol{\lambda}_i^*\|^2
+
\sum_{j=1}^{d_i}
\log
\left[
1-\Phi
\left\{
\alpha_j(z_{i,1},\ldots,z_{i,j-1})
-\lambda_{i,j}^*
\right\}
\right]
\right\}
\]
Therefore, the MET estimator of the censored probability is
\[
\widehat{p}_i
=
\frac{1}{M}
\sum_{m=1}^{M}
w(\mathbf{z}_i^{(m)})
=
\frac{1}{M}
\sum_{m=1}^{M}
\exp
\left\{
\psi_i(\mathbf{z}_i^{(m)},\boldsymbol{\lambda}_i^*)
\right\}
\]

\section{Sensitivity to the maximum number of censored nearest neighbors}
\label{app:neighbor_sensitivity}

\begin{figure}[!htbp]
    \centering

    \begin{subfigure}{0.95\textwidth}
        \centering
        \includegraphics[width=\textwidth]{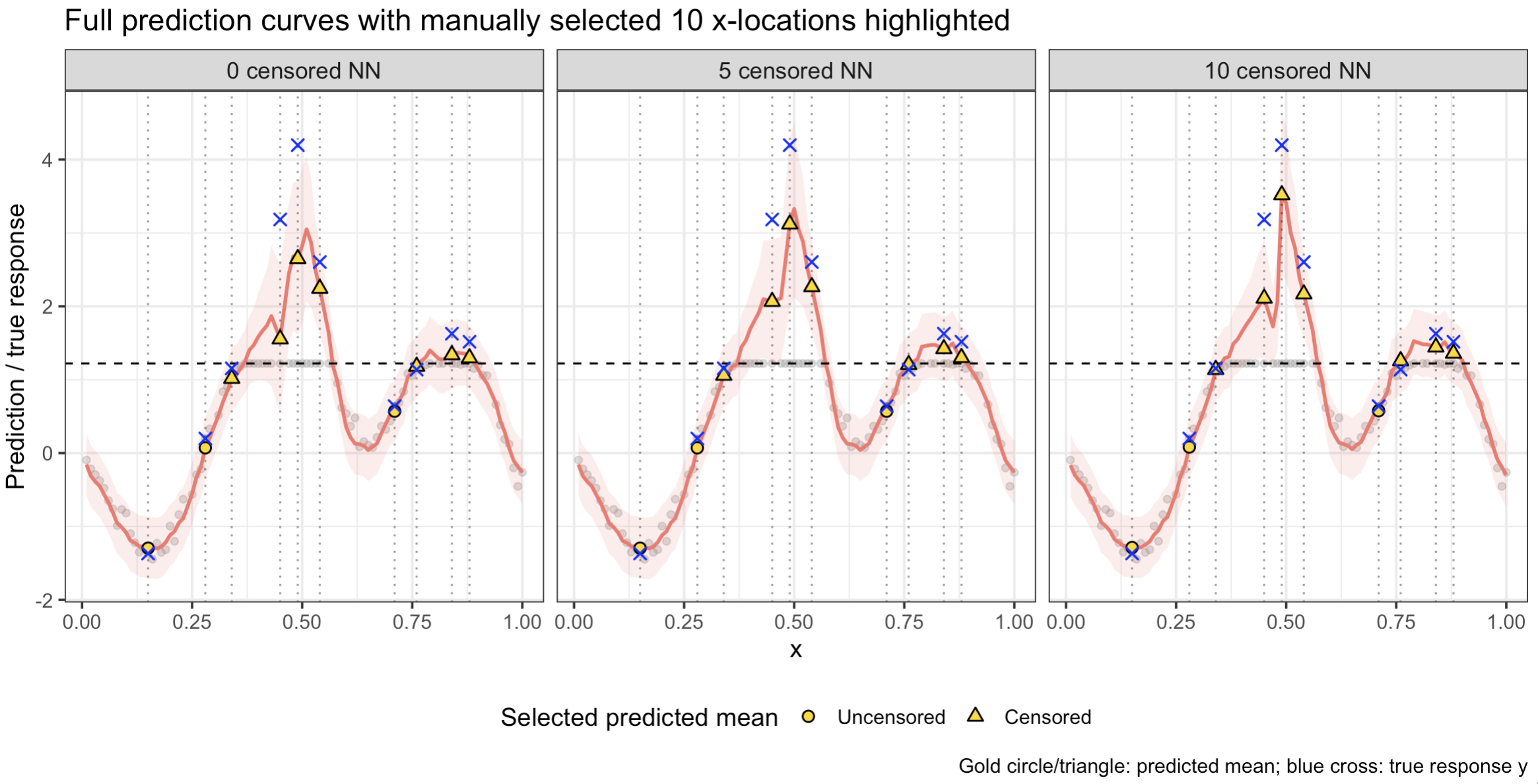}
        \caption{Fitted prediction curves, 95\% predictive intervals, selected prediction locations, and true responses under different maximum numbers of censored nearest neighbors.}
        \label{fig:sensitivity_full}
    \end{subfigure}

    \vspace{0.2cm}

    \begin{subfigure}{0.95\textwidth}
        \centering
        \includegraphics[width=\textwidth]{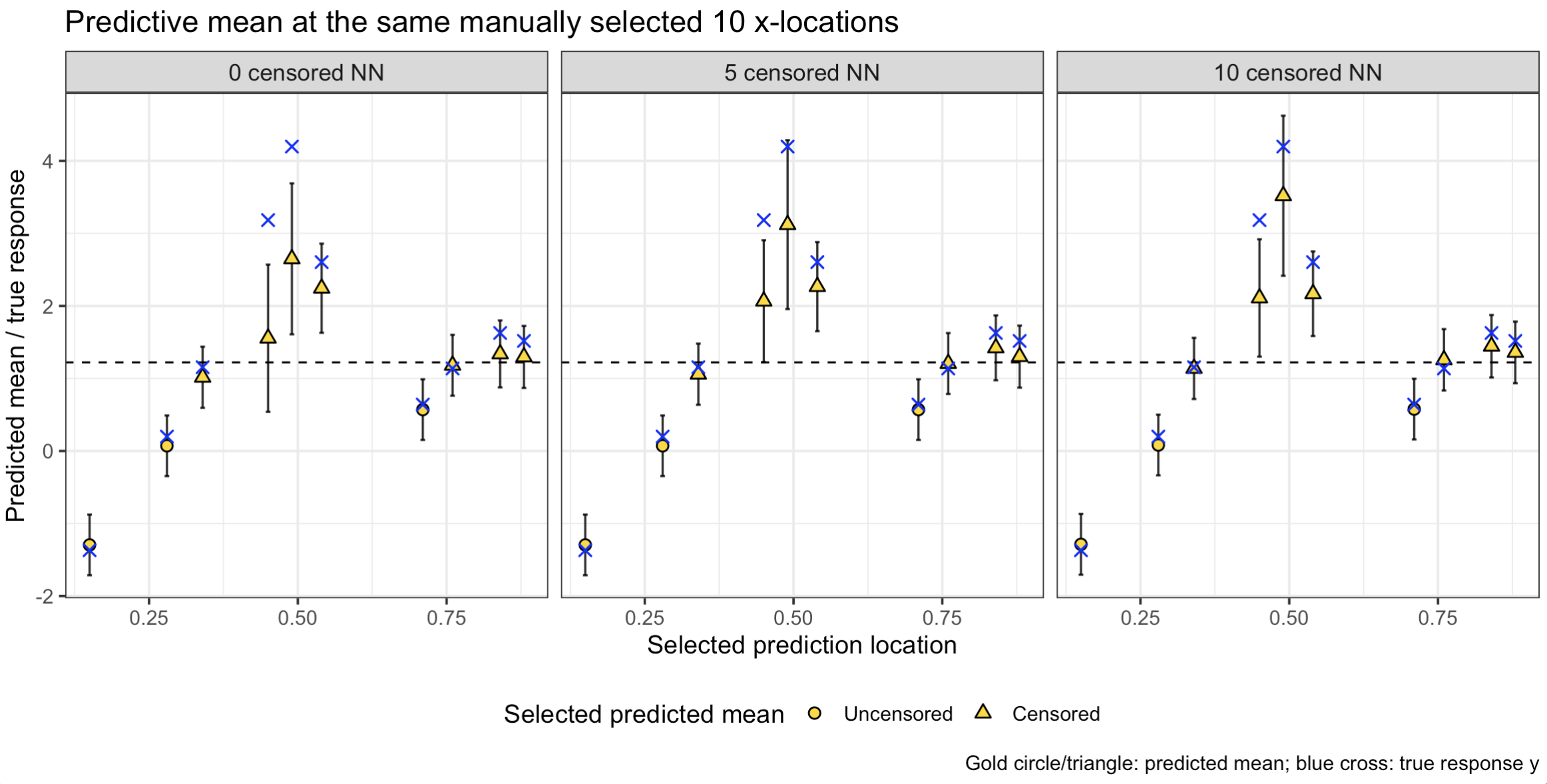}
        \caption{Predicted means, 95\% predictive intervals, and true responses at the same ten selected prediction locations.}
        \label{fig:sensitivity_mean}
    \end{subfigure}

    \caption{Sensitivity analysis for different maximum numbers of censored nearest neighbors in the one-dimensional example.}
    \label{fig:sensitivity_neighbor_rule}
\end{figure}

This appendix reports a sensitivity analysis for the maximum number of censored nearest neighbors allowed in the prediction neighborhood. We compare three rules: no censored nearest neighbors, at most five censored nearest neighbors, and at most ten censored nearest neighbors. The results are shown in Figure~\ref{fig:sensitivity_neighbor_rule} and Table~\ref{tab:censored_neighbor_sensitivity}.

\begin{table}[htbp]
\centering
\caption{Sensitivity analysis for different maximum numbers of censored nearest neighbors in the one-dimensional example.}
\label{tab:censored_neighbor_sensitivity}
\begin{tabular}{lccccc}
\hline
Neighbor rule & MSPE & MAE & CRPS & Coverage & AIL \\
\hline
\multicolumn{6}{c}{\textit{Uncensored region}} \\
\cline{1-6}
No censored NN         & \textbf{0.0369} & \textbf{0.1539} & \textbf{0.1092} & 0.9809 & 0.8388 \\
At most 5 censored NN  & 0.0374 & 0.1543 & 0.1098 & 0.9809 & 0.8389 \\
At most 10 censored NN & 0.0382 & 0.1559 & 0.1110 & 0.9745 & \textbf{0.8384} \\
\hline
\multicolumn{6}{c}{\textit{Censored region}} \\
\cline{1-6}
No censored NN         & 0.3595 & 0.4421 & 0.3216 & 0.7935 & 1.1820 \\
At most 5 censored NN  & \textbf{0.2588} & \textbf{0.3700} & \textbf{0.2638} & \textbf{0.8839} & 1.1439 \\
At most 10 censored NN & 0.3075 & 0.3779 & 0.2804 & 0.8452 & \textbf{1.0675} \\
\hline
\end{tabular}
\begin{flushleft}
\footnotesize
\textcolor{black}{Note: NN = nearest neighbors; AIL = average interval length. Bold values indicate the best value within each region.}
\end{flushleft}
\end{table}

\newpage

\section{Sensitivity analysis for unbalanced replication and larger noise variance}
\label{app:unbalanced_noise}

To further assess the robustness of the proposed method, we conducted an additional one-dimensional sensitivity analysis under highly unbalanced replication and larger noise variance. This analysis is designed to address settings in which some input locations are tested many times while others are tested only once, which is common in industrial experiments.

Table~\ref{tab:unbalanced_settings} summarizes the three simulation scenarios. Scenario 1 is a balanced sparse replication setting, where each of the 300 unique input locations has one replicate. Scenario 2 considers highly unbalanced replication, where some input locations have one replicate and others have ten replicates, increasing the total number of observations to 1246 while keeping the number of unique input locations fixed at 300. Scenario 3 uses the same unbalanced replication pattern as Scenario 2 but increases the noise variance from \(\sigma^2=0.04\) to \(\sigma^2=0.36\).

\begin{table}[htbp]
\centering
\caption{Simulation settings for evaluating unbalanced repeated measurements and larger noise variance.}
\label{tab:unbalanced_settings}
\begin{tabular}{lccccc}
\toprule
Scenario 
& Replication pattern 
& $n_{\text{unique}}$ 
& $n_{\text{obs}}$ 
& $\sigma^2$ 
& Time (sec) \\
\midrule
1 
& $r_i = 1$ 
& 300 
& 300 
& 0.04 
& 12 \\

2 
& $r_i = 1$ or $10$ 
& 300 
& 1246 
& 0.04 
& 73 \\

3 
& $r_i = 1$ or $10$ 
& 300 
& 1246 
& 0.36 
& 74 \\
\bottomrule
\end{tabular}
\end{table}

The prediction results are reported in Table~\ref{tab:unbalanced_performance}. Comparing Scenario 1 and Scenario 2, the prediction performance in the uncensored region remains similar under unbalanced replication. In the censored region, performance improves under unbalanced replication, with MSPE decreasing from 0.7073 to 0.5234 and CRPS decreasing from 0.4773 to 0.4159. These results suggest that additional replicates at some input locations provide useful information for prediction, even when the replication pattern is uneven.

When the noise variance is increased in Scenario 3, prediction becomes more difficult, as expected. The point prediction errors increase and the predictive intervals become wider. For example, the average interval length increases from 0.7896 to 2.5008 in the uncensored region and from 1.3275 to 4.2469 in the censored region. However, the wider intervals reflect the increased uncertainty, and the empirical coverage remains reasonable, especially in the censored region.

\begin{table}[htbp]
\centering
\caption{Prediction performance under balanced sparse replication, unbalanced replication, and larger noise variance in the one-dimensional example.}
\label{tab:unbalanced_performance}
\begin{tabular}{lccccc}
\toprule
Scenario 
& MSPE 
& MAE 
& CRPS 
& Coverage 
& AIL \\
\midrule
\multicolumn{6}{c}{\textit{Uncensored region}} \\
\midrule
1
& 0.0356 & 0.1486 & 0.1059 & 0.9655 & 0.7607 \\

2
& 0.0418 & 0.1633 & 0.1156 & 0.9459 & 0.7896 \\

3
& 0.4063 & 0.5092 & 0.3588 & 0.9662 & 2.5008 \\
\midrule
\multicolumn{6}{c}{\textit{Censored region}} \\
\midrule
1
& 0.7073 & 0.5776 & 0.4773 & 0.6250 & 1.0967 \\

2
& 0.5234 & 0.5589 & 0.4159 & 0.6842 & 1.3275 \\

3
& 0.6693 & 0.6410 & 0.4711 & 0.9665 & 4.2469 \\
\bottomrule
\end{tabular}
\end{table}

\begin{figure}[!htbp]
    \centering
    \begin{subfigure}{0.95\textwidth}
        \centering
        \includegraphics[width=\textwidth]{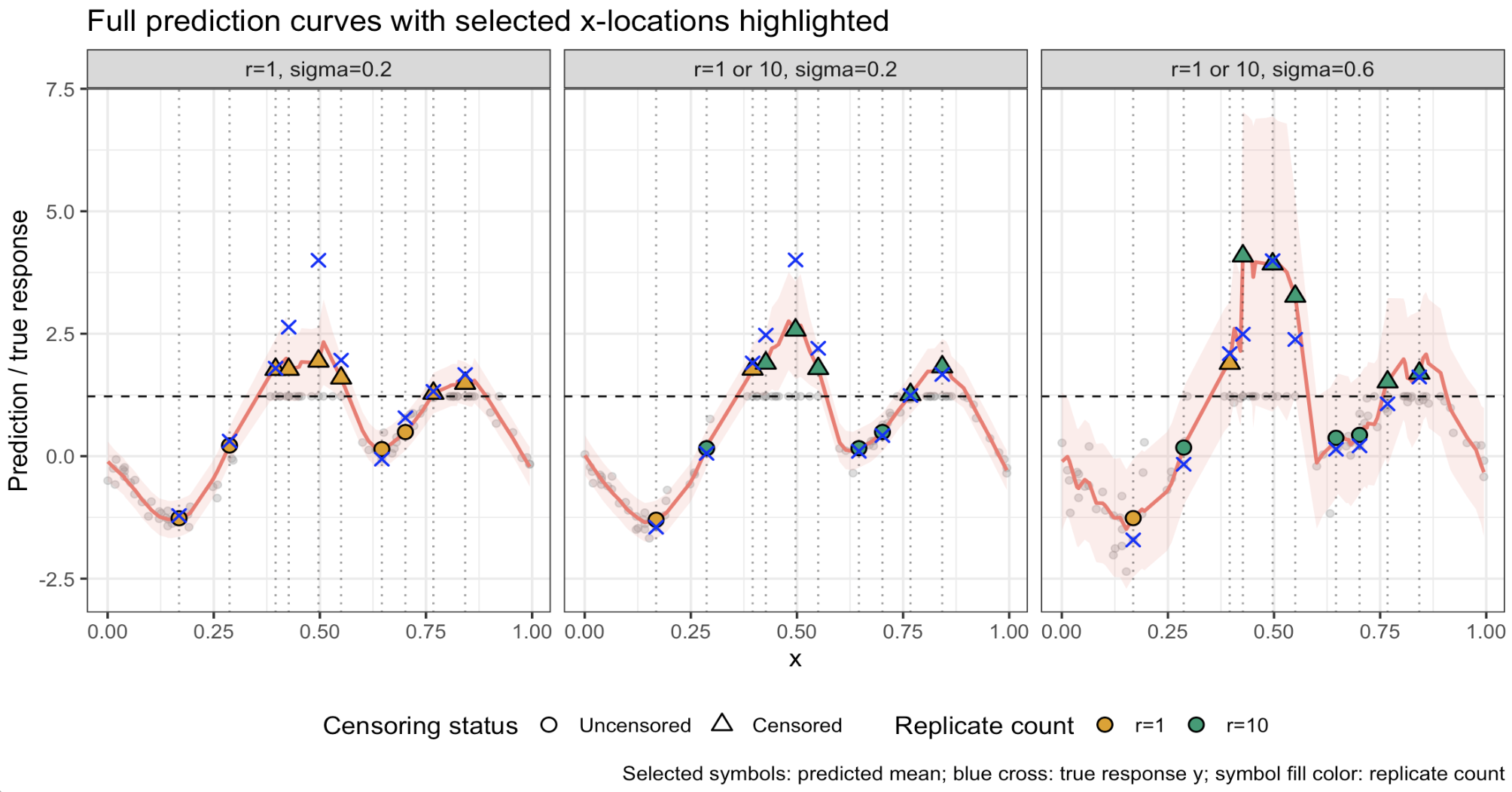}
        \caption{Prediction curves under balanced replication, unbalanced replication, and unbalanced replication with larger noise. The red curve represents the predicted mean, the shaded region represents the 95\% predictive interval, the dashed horizontal line denotes the censoring threshold, and the dotted vertical lines indicate the selected prediction locations. The selected symbols show the predicted means, with shape indicating censoring status and fill color indicating the replicate count. Blue crosses denote the true responses.}
        \label{fig:replication_full}
    \end{subfigure}
    \vspace{0.1cm}
    \begin{subfigure}{0.95\textwidth}
        \centering
        \includegraphics[width=\textwidth]{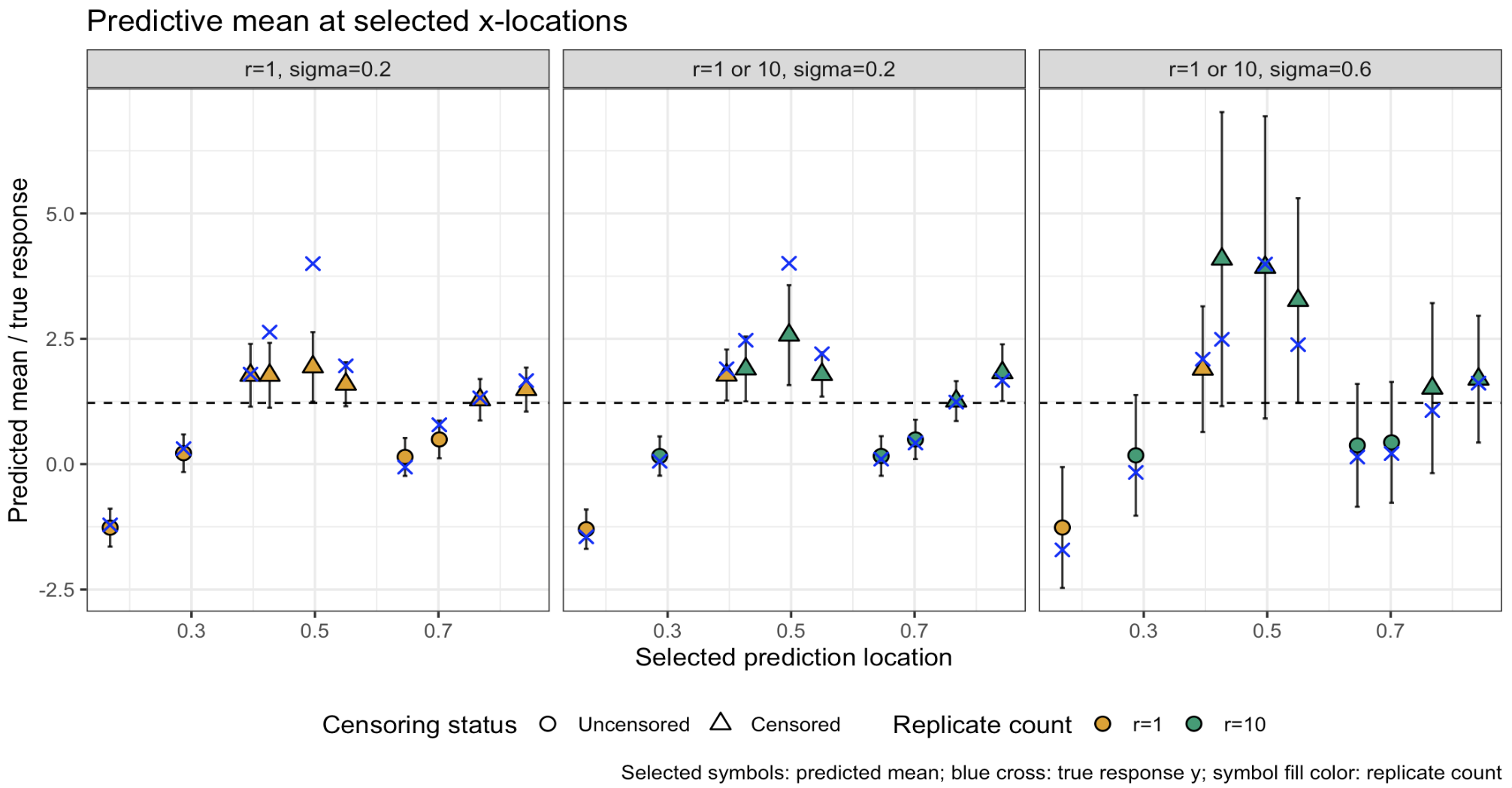}
        \caption{Predicted means, 95\% predictive intervals, and true responses at the same selected prediction locations.}
        \label{fig:replication_selected}
    \end{subfigure}
    \caption{Prediction performance under balanced replication, unbalanced replication, and increased noise variance.}
    \label{fig:replication_noise_prediction}
\end{figure}

Figure~\ref{fig:replication_noise_prediction} provides a visual comparison of the fitted prediction curves, predictive intervals, and selected prediction locations. The proposed method remains stable under highly unbalanced replication. Under larger noise variance, the predictive intervals become wider, reflecting the increased uncertainty in the data-generating process.

\end{document}